\documentclass[sn-mathphys,Numbered]{sn-jnl}% Math and Physical Sciences Reference Style

\usepackage{graphicx}%
\usepackage{multirow}%
\usepackage{amsmath,amssymb,amsfonts}%
\usepackage{amsthm}%
\usepackage{mathrsfs}%
\usepackage[title]{appendix}%
\usepackage{xcolor}%
\usepackage{textcomp}%
\usepackage{manyfoot}%
\usepackage{booktabs}%
\usepackage{algorithm}%
\usepackage{algorithmicx}%
\usepackage{algpseudocode}%
\usepackage{listings}%
\usepackage{soul}
\usepackage{ulem}
\theoremstyle{thmstyleone}%
\theoremstyle{thmstyletwo}%
\theoremstyle{thmstylethree}%

\begin{document}

\title{On the temporal tweezing of cavity solitons}

\author[1]{\fnm{Julia} \sur{Rossi}} %\email{julia.rossi@gmail.com}

\author[2]{\fnm{Sathyanarayanan} 
%\spfx{G.} 
\sur{Chandramouli}} %\email{kevrekid@umass.edu}

\author[1]{\fnm{Ricardo} \sur{Carretero-Gonz\'alez}} %\email{rcarretero@sdsu.edu}

\author[2]{\fnm{Panayotis} \spfx{G.} \sur{Kevrekidis}} %\email{kevrekid@umass.edu}

\affil[1]{\orgdiv{Nonlinear Dynamical Systems Group\footnote{{\tt URL:} http://nlds.sdsu.edu/},
Computational Science Research Center\footnote{{\tt URL:} http://www.csrc.sdsu.edu/}, and
Department of Mathematics and Statistics},
\orgname{San Diego State University}, \orgaddress{\city{San Diego}, \postcode{92182-7720}, \state{California}, \country{USA}}}

\affil[2]{\orgdiv{Department of Mathematics and Statistics}, 
\orgname{University of Massachusetts}, \orgaddress{\city{Amherst}, \postcode{01003-4515}, \state{Massachusetts}, \country{USA}}}

\abstract{

%Ref.~\cite{tweeze}
Motivated by the work of J.K.~Jang et al., Nat.~Commun.~{\bf 6}, 7370 (2015), 
where the authors experimentally tweeze cavity solitons in a passive loop of optical 
fiber, we study the amenability to tweezing of cavity solitons as the properties of 
a localized tweezer are varied.
The system is modeled by the Lugiato-Lefever equation,
a variant of the complex Ginzburg-Landau equation.
We produce an effective, localized, trapping tweezer potential by 
assuming a Gaussian phase-modulation of the holding beam.
The potential for tweezing is then assessed as the total (temporal) 
displacement and speed of the tweezer are varied, and corresponding
phase diagrams are presented.
As the relative speed of the tweezer is increased we find two
possible dynamical scenarios:
successful tweezing and release of the cavity soliton.
We also deploy a non-conservative variational approximation (NCVA) 
based on a Lagrangian description which reduces the 
original dissipative partial differential equation to a set of 
coupled ordinary differential equations for the cavity soliton 
parameters. We illustrate the ability of the NCVA to
accurately predict the separatrix between successful and failed
tweezing. This showcases the versatility of the NCVA to provide 
a low-dimensional description 
of the experimental realization of the temporal tweezing.
}

\keywords{Optical tweezers, cavity solitons, complex Ginzburg-Landau equation,
Lugiato-Lefever equation, non-conservative Lagrangian formulation}

\maketitle

\bmhead{Acknowledgments}

R.C.G.~gratefully acknowledges support from the US National Science Foundation
under Grants PHY-1603058 and PHY-2110038.
P.G.K.~gratefully acknowledges support from the US National Science Foundation
under Grants DMS-2204702 and PHY-2110030.

%\section*{Declarations}
%
%Some journals require declarations to be submitted in a standardized format. Please check the Instructions for Authors of the journal to which you are submitting to see if you need to complete this section. If yes, your manuscript must contain the following sections under the heading `Declarations':
%
%\begin{itemize}
%\item Conflict of interest/Competing interests (check journal-specific guidelines for which heading to use)
%\item Ethics approval 
%\item Consent for publication
%\item Availability of data and materials
%\item Authors' contributions
%\end{itemize}

%%%%%%%%%%%%%%%%%%%%%%%%%%%%%%%%%%%%%%%%%%%%%%%%%%%%%%%%%%%%%%%%%
\section{Introduction}
%%%%%%%%%%%%%%%%%%%%%%%%%%%%%%%%%%%%%%%%%%%%%%%%%%%%%%%%%%%%%%%%%

An optical tweezer, i.e.,~a single-beam gradient force trap, can capture, 
manipulate and move nanometer and micron-sized dielectric particles in 
space using a highly focused laser beam~\cite{Ashkin1970,Ashkin1986}. 
Optical tweezers have been used in physics and biology to manipulate 
objects and measure forces~\cite{ChuOpt}. 
The optical trapping and manipulation of the temporal position of light 
pulses is highly desirable as it may have direct implications for optical 
information processing. In this case, information is treated as a 
sequence of pulses that can be stored and reconfigured by trapping 
ultrashort pulses of light and dynamically moving them around in time. 
In particular, a temporal tweezer can exert similar control over 
ultrashort light pulses in time as reported in Ref.~\cite{tweeze}. 
The optical trapping and manipulation of light pulses 
is useful in optical information processing~\cite{info1,info2,info3,info4}
where the information is represented as a sequence of pulses. 
Temporal tweezing is an effective method to trap ultrashort pulses of 
light and move them around in {\it time} in order to store and reconfigure 
the information. Optical information processing is partly achieved in 
slow-light~\cite{info5,info6,info7,info8} and nonlinear cross-phase 
modulation effects~\cite{info9,info10,info11,info12,info13,info14}, 
however neither approach allows for the independent control of light 
pulses within the sequence. It should be noted that 
solitary wave transfer in continuum and discrete (gain and loss)
media has been previously illustrated, e.g.,
in Ref.~\cite{PhysRevE.66.015601},
while the use of potentials (such as time-dependent
optical~\cite{rcg25,rcg34} or Bessel~\cite{PhysRevA.76.053601} lattices) 
has been used to drive solitary waves in atomic and optical systems. 
Such ideas of capture, release and overall control of the solitary waves 
have been popular in other areas too, including, e.g., in electrical
lattice experiments~\cite{lars}.
Based on Ref.~\cite{tweeze}, we present a detailed analysis of 
temporal tweezing by means of direct numerical simulations as well as 
an analytical low-dimensional reduction of the wave's
dynamics based on a non-conservative variational 
approach (NCVA)~\cite{JuliaNCVA}.

We investigate temporal tweezing of cavity solitons in a passive loop of 
optical fiber pumped by a continuous-wave laser beam which is described by a 
modified Lugiato-Lefever (LL) partial differential equation (PDE) model
which, in turn, is a special case of the celebrated complex Ginzburg-Landau
equation~\cite{CGL,CGL2}. 
In the experiments described in Ref.~\cite{tweeze}, 
temporal cavity solitons (CSs) were created as picosecond 
pulses of light that recirculate in a loop of optical fiber and are 
exposed to temporal controls in the form of a gigahertz phase modulation. 
It has been shown, both theoretically and experimentally, that the CSs 
are attracted and trapped to phase maxima, suppressing all soliton 
interactions. These trapped CSs can then be manipulated in time, 
either forward or backward, which is known as temporal tweezing. 
We study the existence and dynamics of temporally tweezed CSs. 
The key phenomena reported herein are parametric regions
separating the following tweezing scenarios:
(i) successful temporal tweezing of the CS where the CS moves 
as prescribed by the tweezer and
(ii) the failed tweezing regime where the effective potential moves too fast and leaves the CS behind.
We also apply the NCVA to reduce the LL dynamics of CSs to a set
of ordinary differential equations (ODEs) on the CS parameter. 
We therefore, identify regions of temporal tweezing, and 
compare to the full numerical solutions of the original LL PDE. 
We find very good agreement between the two for the 
parametric regime considered, illustrating the ability 
of this effective reduction to provide an accurate 
characterization of the dynamics.

The manuscript is organized as follows. 
In Sec.~\ref{section:TTweeze} we introduce the temporal tweezing approach suggested 
in Ref.~\cite{tweeze} and decribe the LL model. 
We then add a Gaussian phase-modulation to the LL model and simulate the moving and manipulation of the CS. 
In Sec.~\ref{section:TweezeNCVA} we provide a brief description of the NCVA 
approach and its formulation within the LL model. Then, the rest of the section
is devoted to the application of the NCVA to capture the tweezing of a CS. 
In Sec.~\ref{section:TweezeResults} we identify parameter regimes for the existence 
of trapped CSs and follow the tweezing dynamics from the ensuing manipulation of 
the phase-modulation of a continuous-wave holding beam. 
Finally, in Sec.~\ref{secConclusion} we summarize our key findings and 
provide possible avenues for future research.

%%%%%%%%%%%%%%%%%%%%%%%%%%%%%%%%%%%%%%%%%%%%%%%%%%%%%%%%%%%%%%%%%
\section{The Full Model: Lugiato-Lefever Equation and Temporal Phase Modulation}
\label{section:TTweeze}
%%%%%%%%%%%%%%%%%%%%%%%%%%%%%%%%%%%%%%%%%%%%%%%%%%%%%%%%%%%%%%%%%

In our analysis of temporal tweezing we begin with dissipative solitons in 
externally-driven nonlinear passive cavities, the so-called temporal 
CSs~\cite{XuCoenRef22a,XuCoenRef22b,info15,info17,info19,info21}. 
In a passive loop of optical fiber these light pulses can persist 
without losing shape because the dispersive temporal spreading is balanced 
by the material nonlinearity. Also, CSs persist without losing their 
intensity by drawing power from a continuous-wave (cw) ``holding'' 
laser beam driving the cavity. Multiple CSs may be simultaneously present 
in the optical loop and positioned independently temporally~\cite{info16}. 
Here, we report on the trapping of CSs and the dynamical manipulation 
through selectively altering the phase profile of the holding beam. 

%%%%%%%%%%%%%%%%%%%%%%%%%%%%%%%%%%%%%%%%%%%%%%%%%%%%%%%%%%%%%%%%%
\subsection{Theory of Temporal Tweezing} 
\label{secTweezeTheory}
%%%%%%%%%%%%%%%%%%%%%%%%%%%%%%%%%%%%%%%%%%%%%%%%%%%%%%%%%%%%%%%%%

Temporal tweezing requires a CS with an attractive time-domain drift towards the 
maxima of the intracavity phase profile. The attraction is due to the CSs shifting 
their instantaneous frequencies in response to a phase modulation. 
The gain and loss mechanisms inherent to this system may be captured by a variant
of the celebrated complex Ginzburg-Landau (cGL) equation~\cite{CGL,CGL2}.
In particular, following Refs.~\cite{tweeze,LL,LLE}, we will model the system under 
consideration by the following Lugiato-Lefever (LL) equation:
\begin{eqnarray}
z_R \frac{\partial E }{\partial z} = \left[ -\alpha - i
{\delta} - i L \frac{\beta_2}{2} \frac{\partial^2}{\partial \tau^2} + i \gamma L |E|^2 \right] E + \sqrt{\theta}E_{\rm in},
\label{eq:LLETweeze1}
\end{eqnarray}
where $z$ is the slow time describing the intracavity field envelope $E(z,\tau)$ 
and $\tau$ is the fast time describing the 
temporal profile of the field envelope in a reference frame traveling at the 
group velocity of the holding beam in the cavity. The cavity roundtrip time 
is $z_R$ and the field of the holding beam is $E_{\rm in}$ with power 
$P_{\rm in} = |E_{\rm in}|^2$. The cavity losses are accounted by 
$\alpha = \pi/\mathscr{F}$, where $\mathscr{F}$ is the cavity finesse.
The phase detuning of the intracavity field to the closest cavity 
resonance of order $l$ is given by ${\delta} = 2\pi l - \phi_0$ where $\phi_0$ 
is a linear phase-shift over one roundtrip with respect to the holding beam. 
Finally, $L$ is the cavity length, $\beta_2$ is the dispersion coefficient of 
the fiber, $\gamma$ is the nonlinear coefficient of the fiber, and $\theta$ is 
the input coupler power transmission coefficient. 

In what follows, for ease of exposition, we will work in adimensional units. 
For that purpose, the LL Eq.~(\ref{eq:LLETweeze1}) may be adimensionalized 
by introducing a dimensionless slow time $z' = \alpha z / z_R$ and 
a dimensionless fast time $\tau' = \tau \sqrt{2\alpha /(L |\beta_2|)}$. 
We also use a dimensionless complex field amplitude 
$v(z',\tau') = E(z,\tau) \sqrt{\gamma L/\alpha}$ and a dimensionless holding 
beam $v_{\rm in} = E_{\rm in}\sqrt{\gamma L \theta /\alpha^3}$. 
For convenience, we drop the primes in the notation of $z'$ and $\tau'$, such 
that Eq.~(\ref{eq:LLETweeze1}) becomes the dimensionless mean-field LL equation 
\begin{eqnarray}
\frac{\partial v }{\partial z} = -(1+i \Delta) v + i |v|^2 v 
+i  \frac{\partial^2 v }{\partial \tau^2} + v_{\rm in},
\label{eq:dimensionlessLLE}
\end{eqnarray}
where $\Delta = \delta/\alpha$ is the effective dimensionless detuning and we have chosen ${\rm sgn}(\beta_2)=-1$. 

Homogeneous and steady ($\tau$- and $z$-independent) states, $v(z,\tau)=v_s$, 
of Eq.~(\ref{eq:dimensionlessLLE}), where the CSs will be supported, satisfy 
\begin{eqnarray}
|v_s|^2 v_s = -i v_s + \Delta v_s+ i v_{\rm in}. 
\label{steadyStateLL}
\end{eqnarray}
Defining the intracavity background field intensity 
$I_s \equiv |v_s|^2$, one can write 
the steady state solution in implicit form
\begin{eqnarray}
\nonumber
v_s = \frac{v_{\rm in}}{1+ i (\Delta - I_s)},
\end{eqnarray}
which displays the dependence on the detuning $\Delta$ and holding beam power $v_{\rm in}$. 
{This implicit relationship may be cast in the form of a well-known cubic equation for dispersive optical 
bistability~\cite{Grelu,LL,info2,Gomila2007}}
\begin{eqnarray}
I_s^3 - 2\Delta I_s^2 + (1+ \Delta^2) I_s = I_0,
\label{steadystate2}
\end{eqnarray}
where
$I_0 \equiv |v_{\rm in}|^2$ is the holding beam intensity. 
For small detuning, $\Delta < \sqrt{3}$, Eq.~(\ref{steadystate2}) has only one 
solution for the the steady state $I_s$ given a specific holding beam power 
$v_{\rm in}$. For large detuning, $\Delta > \sqrt{3}$, there are three 
solutions for $I_s$ given a specific holding beam power $v_{\rm in}$. 
The homogeneous solution is bistable since two solutions are stable while 
the other solution is unstable. 
%The transition between the states occurs 
%through a pitchfork bifurcation.

We follow the approach of Refs.~\cite{tweeze,firth96} to study the effect 
of phase-modulation of the holding field. The phase-modulation is imprinted 
into a constant holding beam which creates an effective potential necessary 
to attract, trap, and manipulate a CS. We assume a phase-modulation 
temporal profile $\phi(\tau)$ and rewrite the holding beam using 
\begin{eqnarray}
\label{uin}
v_{\rm in} (\tau) = u_{\rm in} \exp[i \phi(\tau)], 
\end{eqnarray}
where $u_{\rm in}$ is a constant scalar (whose square corresponds to
the intensity of the holding beam). Substituting Eq.~(\ref{uin}) 
into Eq.~(\ref{eq:dimensionlessLLE}) with the ansatz 
$v(z, \tau) = u(z, \tau) \exp[i \phi(\tau)]$ yields 
\begin{eqnarray}
i u_z + |u|^2 u + u_{\tau\tau} -& (\Delta + (\phi')^2) u + 2i u_{\tau} \phi' = 
- i (1+\phi'') u + i u_{\rm in},
\label{eq:LLETweeze}
\end{eqnarray}
where primes denote derivatives with respect to $\tau$.
The cw intracavity field on which the CSs are supported has the 
same phase modulation as that imposed on the external holding beam in 
this non-dimensional form. In this form,
the phase modulation of the holding beam introduces the following
terms affecting the dynamics of the field envelope:
(i) $(\phi')^2$ acts like an effective potential caused by the 
phase modulation,
(ii) the term $ 2i u_{\tau} \phi'$ describes the effect of drift through the
gradient term $u_{\tau}$ where $2 \phi'$ represents a drift ``speed''~\cite{ParraRivas2014},
and (iii) the $\phi''$ term induces an additional loss in the system caused by 
the phase modulation.

Steady state solutions of the LL Eq.~(\ref{eq:LLETweeze}) are subject to 
an additional constraint stemming from the balance condition $dP/dz= 0$, where 
\begin{eqnarray}
\nonumber
P =\int_{-\infty}^{\infty} |u|^2 d \tau,
\end{eqnarray}
is the total power (mathematically, the squared $L^2$ norm) of the cavity solitons.
The evolution of $P$ can be found by multiplying Eq.~(\ref{eq:LLETweeze}) by $u^*$, as well as
the complex conjugate of Eq.~(\ref{eq:LLETweeze}) by $u$, and then adding and 
integrating the resulting equations~\cite{Theocharis2006}. 
For the LL Eq.~(\ref{eq:LLETweeze}), it is straightforward to find the following 
constraint condition for a steady state solution:
\begin{eqnarray}
\int_{-\infty}^{\infty} \Big( - |u|^2 - \phi'' |u|^2 
- \phi' \left(u_{\tau} u^* + u_{\tau}^* u \right) +
% \frac{1}{2} (u^* + u ) 
\mathrm{Re} (u)
u_{\rm in} \Big) d \tau = 0,
%\label{LLConstraint1}
%\\&\implies \int_{-\infty}^{\infty} \Big(-|u|^2-(\phi^{\prime}|u|^2)_{\tau}+{\rm Re}(u) u_{\rm in}\Big) d{\tau}=0,
\end{eqnarray} 
which can be rewritten as
\begin{eqnarray}
\int_{-\infty}^{\infty} \Big(-|u|^2+{\rm Re}(u) u_{\rm in}\Big) d{\tau}=0,
\label{LLConstraint}
\end{eqnarray} 
which demonstrates the {same}  power balance constraint, as expected, for (non)-tweezed cavity solitons. 
This power-balance constraint will need to be enforced in the system, and, in turn, 
will fix the homogeneous background pedestal $v_s$, upon choosing the detuning $\Delta$ and holding beam $u_{\rm in}$ parameters. 

%%%%%%%%%%%%%%%%%%%%%%%%%%%%%%%%%%%%%%%%%%%%%%%%%%%%%%%%%%%%%%%%%
\subsection{{Temporal} tweezing of Cavity Solitons} 
\label{section:TweezePDE}
%%%%%%%%%%%%%%%%%%%%%%%%%%%%%%%%%%%%%%%%%%%%%%%%%%%%%%%%%%%%%%%%%

Our analysis involves temporal cavity solitons described by LL Eq.~(\ref{eq:LLETweeze}) 
stored in a passive loop of optical fiber pumped by a cw laser beam. The {tweezed} CS is 
trapped into a specific time slot through phase-modulation of the holding beam, 
and moved around in time by manipulating the phase profile. Experimentally, 
a modulator imprints a time-varying electric signal $\phi(\tau)$ into the phase 
of the cw holding laser driving the cavity. For the purpose of the LL 
Eq.~(\ref{eq:LLETweeze}), we used a ``natural'' localized phase modulation
with a Gaussian profile of the form
\begin{eqnarray}
\phi(\tau) = h_{\phi} \exp\left( -\frac{(\tau - \tau_0)^2}{2 \sigma_{\phi}^2} \right),
\label{phi}
\end{eqnarray}
where $h_{\phi}$, $\sigma_{\phi}$, and $\tau_0$ describe, respectively, the height, 
width, and center position of the phase profile. 
%
%For the LL model, the first and 
%second derivative of the phase profile are, respectively, 
%.
%\begin{eqnarray}
%\phi' &= \frac{d \phi}{d \tau} = -\frac{h_{\phi} ( \tau - \tau_0)}{\sigma_{\phi}^2} e^{\left( -\frac{(\tau - \tau_0)^2}{2 \sigma_{\phi}^2} \right)}\label{firstphi}, \\
%\phi'' &= \frac{d^2 \phi}{d \tau^2} = -\frac{h_{\phi} }{\sigma_{\phi}^2} e^{ \left[ -\frac{(\tau - \tau_0)^2}{2 \sigma_{\phi}^2} \right)} \left(1 +\frac{( \tau - \tau_0)^2}{\sigma_{\phi}^2} \right]. \label{secondphi} 
%.
%\end{eqnarray}

In what follows, we will first consider the stationary solutions of the LL model 
in the form $u(z, \tau) = u_0(\tau)$ which are governed by the following ODE:
\begin{eqnarray}
%u_{0,\tau\tau} + \left( |u_0|^2 -\Delta - (\phi')^2 \right) u_0 + 2 i \phi' u_{0, \tau} 
u_0'' + \left( |u_0|^2 -\Delta - (\phi')^2 \right) u_0 + {2 i \phi' u_0'} 
= - i (1+\phi'') u_0 + i u_{\rm in}.
\label{LLNoSol}
\end{eqnarray}
%
%\begin{equation}
%\int_{-\infty}^{\infty} \Big[ - (1 + \phi'') |u_0|^2 
%- \phi' \left( u_0' u_0^* + {u_0^*}' u_0 %\right) +
% \mathrm{Re}(u_0) u_{\rm in} \Big] d \tau = 0.
%\label{LLconstraint0}
%\end{equation}
%
%The self-consistency condition in Eq.~\eqref{LLConstraint} relates the particular value of the detuning 
%$\Delta$ to the other parameters (i.e., $u_{\rm in}$ and {$v_s$}).
%Indeed, this parameter is reminiscent of the frequency of the solitary
%waves in the corresponding Hamiltonian, nonlinear Schr{\"o}dinger (NLS) type
%settings~\cite{chap01:ablowitz}. While in such NLS systems, this
%parameter has freely tunable ranges, in the present dissipative system,
%it is determined by the relevant balance condition discussed above. 
%
In practice, in what follows, the homogeneous background pedestal $v_s$ is determined by solving algebraic equation Eq.~\eqref{steadyStateLL} having made a choice for $\Delta$ and $u_{\rm in}$. This in turn fixes the boundary state ($v_s$) for stationary CS states.
Once stationary soliton solutions of the ODE Eq.~(\ref{LLNoSol}) are identified (consistent with this boundary state)
at $\tau_0 = 0$, their 
``{temporal} tweezability'' (i.e., amenability to transfer via tweezers)
is considered by measuring the amount of soliton 
intensity that remains inside and outside of the effective potential 
$(\phi')^2$ as the center location of the tweezer is 
(subsequently) manipulated.

%%%%%%%%%% Fig %%%%%%%%%%%%%%%%%%%%%%%%%%%%%%%%%%%%%%
\begin{figure}[t!]
\centering
\includegraphics[width=8.5cm]{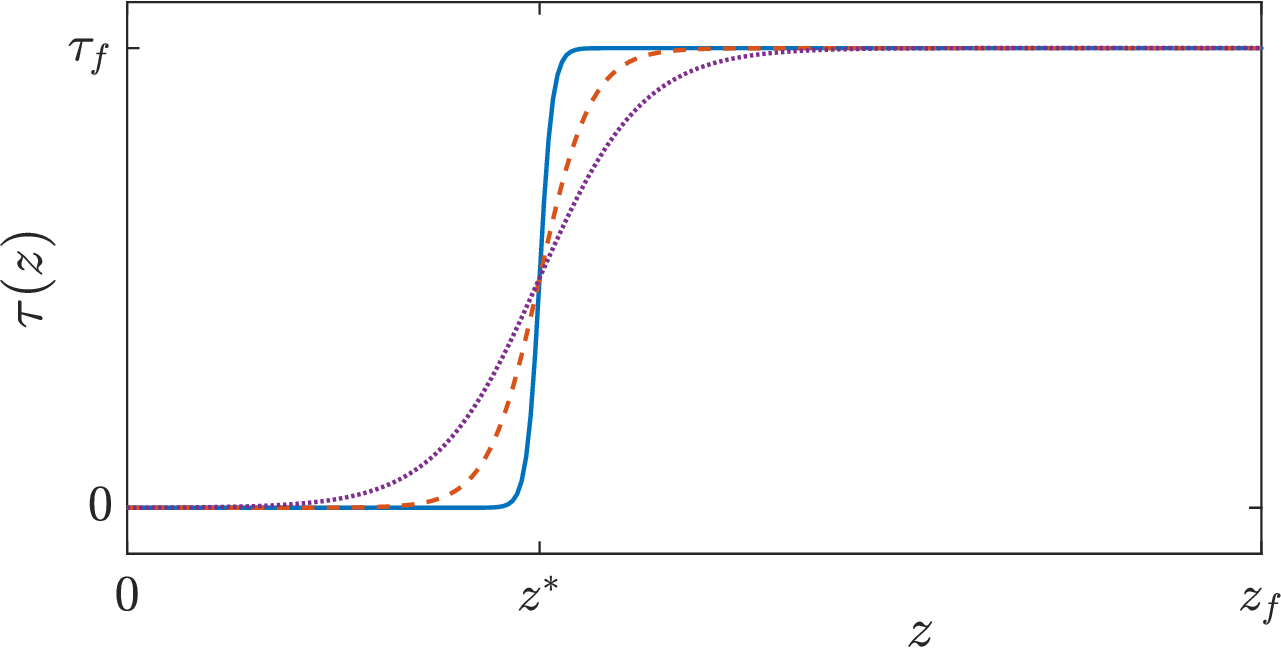}
\caption{
``Trajectory'' of the temporal tweezing. The tweezer starts at $\tau=0$
and ends at $\tau=\tau_f$. The degree of adiabaticity is controlled with the
parameter $\beta$ in Eq.~(\ref{tau0}). The solid (blue) line represents a
fast transition (relatively large $\beta$) which will be 
less likely to tweeze the CS. 
With increasing degrees of adiabaticity (i.e., smaller values of $\beta$), see dashed and
dotted curves, it will be more likely for the tweezing to be successful.}
\label{fig:tau0}
\end{figure}
%%%%%%%%%% Fig %%%%%%%%%%%%%%%%%%%%%%%%%%%%%%%%%%%%%%

To simulate moving and manipulating the cavity soliton through the phase 
profile we consider the following ``motion'' in the evolution variable $z$:
\begin{eqnarray}
\tau_0 (z) = \frac{\tau_f}{2} \left( \frac{\tanh [\beta (z - z^*)]}{\tanh(\beta z^*)}
+ 1\right),
\label{tau0}
\end{eqnarray} 
where $\tau_f$ is the final fast time $\tau$ at which the phase profile 
stops, $z^*$ is the slow time for the phase profile to reach $\tau_f/2$, 
and $\beta$ is the adiabaticity parameter which describes how fast the 
effective potential moves the temporal profile centered at $\tau_0$. 
Figure \ref{fig:tau0} depicts the trajectory of the temporal tweezing.

%In the numerical results Sec.~\ref{section:TweezeResults}, we show {two} 
%domains for the dynamics of the CS in the $(\beta,\tau_f)$ parameter plane: 
%%
%(i) successful temporally tweezed CS (where the CS moves with effective potential)   
%%(\RED{ We now checked that unless there is an underlying instability of the tweezed 
%%soliton state, the dissipated bulk state is not observed. Verified through numerical experiments. }) 
%and (ii) failed temporally tweezed CS for which the effective potential moves too fast 
%and the CS is left behind in the wake of the tweezer, with instead the formation of 
%a small amplitude localized modulation of the background pedestal (due to the presence 
%of the effective potential). We will also refer to this absence of the CS in the 
%tweezer as the state of ``non-tweezing".
%Between these two states, which are present for broad 
%parametric regimes, there exists a relatively 
%narrow separatrix where the process is marginal and
%the CS is only partially trapped.  %(called non-tweezed CS). 
%
%{\bf PGK: do we need here the paragraph: ``In the numerical
%results...''? Not sure what it adds here.}

Having given the PDE setup of our considerations, we
now turn to its effective low-dimensional description
via the non-conservative variational approximation.

%%%%%%%%%%%%%%%%%%%%%%%%%%%%%%%%%%%%%%%%%%%%%%%%%%%%%%%%%%%%%%%%%
\section{Non-conservative Variational Approximation}
\label{section:TweezeNCVA}
%%%%%%%%%%%%%%%%%%%%%%%%%%%%%%%%%%%%%%%%%%%%%%%%%%%%%%%%%%%%%%%%%

%%%%%%%%%%%%%%%%%%%%%%%%%%%%%%%%%%%%%%%%%%%%%%%%%%%%%%%%%%%%%%%%%
\subsection{Preliminaries}
\label{secNCVA:prelim}
%%%%%%%%%%%%%%%%%%%%%%%%%%%%%%%%%%%%%%%%%%%%%%%%%%%%%%%%%%%%%%%%%
%\section{Non-conservative Variational Approximation}

Let us now consider a semi-analytical method to dynamically reduce the LL equation
to an effective ``particle'' picture described by a set of coupled ODEs on the CS
parameters (height, position, width, phase, velocity, and chirp).
We use the so-called non-conservative variational approximation (NCVA) for
PDEs as described in Ref.~\cite{JuliaNCVA}. This is inspired by the
Lagrangian (and Hamiltonian) formulation
of non-conservative mechanical systems originating in the work
of Refs.~\cite{Galley,Galley:14}; see also Ref.~\cite{galley2014principle}.
For completeness, let us briefly describe the NCVA approach (for more details please
consult Ref.~\cite{JuliaNCVA}). To employ the NCVA, we consider two sets of 
coordinates $u_1$ and $u_2$. As proposed by Galley and 
collaborators~\cite{Galley,Galley:14}, the coordinates are fixed at an initial 
time ($z_i$), but are not fixed at the final time ($z_f$). After applying 
variational calculus for a non-conservative system, both paths are set equal, 
$u_1= u_2$, and identified with the physical path $u$, the so-called physical 
limit (PL). The action functional for $u_1$ and $u_2$ is defined as the 
total line integral of the difference of the Lagrangians between the paths 
plus the line integral of the functional ${\cal R}$ which describes the 
generalized non-conservative forces and depends on both paths: 
\begin{eqnarray}
\nonumber
%\hspace{-0.1cm} S = \int_{t_i}^{t_f} dt \left[ L (u_1, \ldots,t) - L(u_2, \ldots,t) + R \right].
S &=& \int_{z_i}^{z_f}  \left(\mathcal{L} (u_1, u_{1,z}, u_{1,\tau}, \ldots,z) 
- \mathcal{L}(u_2, u_{2,z}, u_{2,\tau}, \ldots,z) + \mathcal{R} \right)\, dz,
\end{eqnarray}
where the $z$ and $\tau$ subscripts denote partial derivatives
with respect to these variables.
The above action defines a new total Lagrangian:
\begin{equation}
\nonumber
%\mathcal{L} = L (u_1, u_{1,z}, \ldots,t) - L (u_2, u_{2,z}, \ldots,t) + R,
\mathcal{L}_T \equiv \mathcal{L}_1 - \mathcal{L}_2 + \mathcal{R},
\end{equation}
where the first two terms represent the conservative Lagrangian densities with 
$\mathcal{L}_i \equiv \mathcal{L}(u_i, u_{i,z}, u_{i,\tau}, ...,z)$, 
for $i=1,2$, and $\mathcal{R}$ contains all non-conservative terms.
For convenience, $u_+ = (u_1 + u_2)/2$ and $u_- = u_1 - u_2$ are defined 
in such a way that at the physical limit $u_+  \rightarrow \, u$ and 
$u_-  \rightarrow \, 0$. The equations of motion then yield
%
%The conjugate momenta $p_\pm= \partial L/\partial u_{\mp,z} \, 
%\rightarrow \, p_+= \partial L/\partial u_{-,z}$ 
%at the physical limit and converges to
%\begin{equation}
%p= \partial L/\partial u.
%\end{equation}
%
\begin{equation}
    {\frac{\partial}{\partial z}\left(\frac{\delta \mathcal{L}}{\delta {u_z^{*}}}\right)=\frac{\delta \mathcal{L}}{\delta {u^{*}}}+\left[\frac{\delta \mathcal{R}}{\delta u_{-}^{*}}\right]_{\rm PL}},
\end{equation}
where $\delta$ denotes the Fréchet derivatives. {Through this method, we recover the Euler-Lagrange equation for the conservative 
terms and all non-conservative terms are folded into 
$[{\delta \mathcal{R}}/{\delta u_{-}^{*}}]_{\rm PL}$. 
It is crucial to construct the term $\mathcal{R}$ such that its derivative with 
respect to the difference variable $u_-$ at the physical limit gives 
back the non-conservative or generalized forces. This part concludes
the field-theoretic formulation of the non-conservative problem and
so far no approximation has been utilized. }
{In} what follows, we will proceed to solve the above Euler-Lagrange equations using an appropriate ansatz 
to {\it approximately} describe the CS solutions in this non-conservative 
Lagrangian formulation.
{Towards this end, we wish to study the dynamics of the solution restricted on 
what we will refer to as a ``solitonic manifold'', i.e. $u\equiv \overline{u}(\tau,\vec{p}(z))$, for which the modified Euler-Lagrange equations for the effective Lagrangian 
${\overline{L} = \int_{-\infty}^\infty \overline{\mathcal{L}}\, d\tau}$ yield}
\begin{equation}
%\frac{\partial p}{\partial t} = \frac{\partial \mathcal{L}}{\partial u} \equiv \frac{\partial L}{\partial u} + \left[ \frac{\partial R}{\partial u_- }\right]_{\rm PL}.
{\frac{d}{dz}\left( \frac{\partial \overline{L}}{\partial \dot{\Vec{p}}} \right)-\frac{\partial \overline{L}}{\partial \Vec{p}} -  I_{\Vec{p}} = 0,
}
\label{NCVAODE}
\end{equation}
where the integrals $I_{\Vec{p}}$\, are given by 
$I_{\Vec{p}}\equiv \int_{-\infty}^\infty \left[{\partial \mathcal{\overline{R}}}/{\partial \Vec{p}_- }\right]_{\rm PL} d\tau$ 
and the overline indicates that the equations of motion 
have been evaluated on the approximate variational ansatz.

%%%%%%%%%%%%%%%%%%%%%%%%%%%%%%%%%%%%%%%%%%%%%%%%%%%%%%%%%%%%%%%%%
\subsection{NCVA of Tweezed Cavity Solitons}
% \label{section:TweezeNCVA}
%%%%%%%%%%%%%%%%%%%%%%%%%%%%%%%%%%%%%%%%%%%%%%%%%%%%%%%%%%%%%%%%%

We now describe the use of an approximate variational ansatz to describe the 
statics and dynamics of CSs subjected to the tweezing generated by the external 
terms induced by the phase variations of the holding beam.
In particular, we apply the NCVA approach to Eq.~(\ref{eq:LLETweeze}) to analytically 
identify the tweezability regions in parameter space by following the solutions 
to the corresponding reduced system of ODEs given by 
the Euler-Lagrange Eqs.~(\ref{NCVAODE}),
restricting the original infinite-dimensional dynamics to the low-dimensional
dynamics of the solitonic manifold (governing the evolution of the coherent
structures characteristic features).
%
%\begin{eqnarray}
%\frac{\partial L}{\partial u} - \frac{d}{dt} \left( \frac{\partial L }{\partial \dot{u}} \right) + \int_{-\infty}^{\infty} \left[ \frac{\partial \mathcal{R}}{\partial u_- } \right]_{\rm PL} d\tau = 0.
%\label{tweezeEL}
%\end{eqnarray}
%
The CS solution for the LL model sits on a pedestal (background). Therefore, we construct 
$u = v_s + \bar{u}$ where $\bar{u}$ is the NCVA ansatz and $v_s$ is the homogeneous 
steady-state pedestal solution described in Eq~(\ref{steadyStateLL}). Applying the new 
construction of $u$ into Eq.~(\ref{eq:LLETweeze}) produces the following modified LL 
equation for $\bar{u}$:
%%
%\begin{eqnarray}
%&i \bar{u}_z + |\bar{u}|^2 \bar{u} + \bar{u}_{\tau\tau} - (\Delta + (\phi')^2) \bar{u} + 2i \bar{u}_{\tau} \phi' + 2\bar{u} |v_s|^2 + \nonumber \\
%&2 |\bar{u}|^2 v_s + (v_s)^2 \bar{u}^* + (\bar{u})^2 v_s^* + |v_s|^2 v_s - \Delta v_s = \nonumber \\
% &- i v_s + i u_{\rm in} - i (1+\phi'') \bar{u} + (\phi')^2 v_s - (\phi'')v_s.
%\end{eqnarray}
%%
%This equation, simplified using Eq.~(\ref{steadyStateLL}), yields
%
%\begin{eqnarray}
%&i \bar{u}_z + |\bar{u}|^2 \bar{u} + \bar{u}_{\tau\tau} - (\Delta + (\phi')^2 -2|v_s|^2) \bar{u} + 2i \bar{u}_{\tau} \phi' = \nonumber \\[1.0ex]
 %&- i (1+\phi'') \bar{u} + \left((\phi')^2 - 2 %|\bar{u}|^2- \phi''\right) v_s - (v_s)^2 %\bar{u}^* - (\bar{u})^2 v_s^*,
 %\label{LLNCVA}
%\end{eqnarray}
%
\begin{eqnarray}
    & {i\overline{u}_z+|\overline{u}|^2\overline{u}+\overline{u}_{\tau\tau}-\left(\Delta+(\phi^{\prime})^2-2|v_s|^2\right)\overline{u}+2i\overline{u}_{\tau}\phi^{\prime}+{i\phi^{\prime\prime}}\overline{u}}= \nonumber \\[1.0ex]
    &{-i\overline{u}+(-2|\overline{u}|^2+(\phi^{\prime})^2-{i}\phi^{\prime\prime})v_s-(v_s)^2\overline{u}^{*}-(\overline{u})^2v_s^{*}}
    \label{LLNCVA}
\end{eqnarray}
%\Revone{CORRECT}
%\begin{align}
    %&\Revone{i\overline{u}_z+|\overline{u}|^2\overline{u}+\overline{u}_{\tau\tau}-\left(\Delta+(\phi^{\prime})^2-2|v_s|^2\right)\overline{u}+2i\overline{u}_{\tau}\phi^{\prime}+{i\phi^{\prime\prime}}\overline{u}+2|\overline{u}|^2 v_R}\\\nonumber&\Revone{+v_R(\overline{u})^2=-i\overline{u}-2iv_I|\overline{u}|^2+((\phi^{\prime})^2-{i}\phi^{\prime\prime})v_s-(v_s)^2\overline{u}^{*}+iv_I(\overline{u})^2}
%\end{align}
where we have used Eq.~(\ref{steadyStateLL}) to slightly simplify the resulting equation.
%the complex conjugate is denoted with $^*$. 
The conservative part of the LL equation, namely the left-hand side of Eq.~(\ref{LLNCVA}),
originates from the following Lagrangian density: 
\begin{eqnarray}
\nonumber
\overline{{\mathcal{L}}}_{\rm cons} =& \frac{i}{2} \left(\bar{u}\bar{u}_z^*-\bar{u}^* \bar{u}_z  \right) + |\bar{u}_{\tau}|^2 - \frac{1}{2} |\bar{u}|^4 
+\left( \Delta + (\phi')^2 -2|v_s|^2 \right) |\bar{u}|^2 - i\phi' \left(\bar{u}^* \bar{u}_{\tau} - \bar{u}\bar{u}_{\tau}^* \right).
\end{eqnarray}
%\Revone{CORRECT}
%\Revone{
%\begin{align}
%\overline{{\mathcal{L}}}_{\rm cons} =& \frac{i}{2} \left(\bar{u}\bar{u}_z^*-\bar{u}^* \bar{u}_z  \right) + |\bar{u}_{\tau}|^2 - \frac{1}{2} |\bar{u}|^4 
%+\left( \Delta + (\phi')^2 -2|v_s|^2 \right) |\bar{u}|^2  \\\nonumber
%+&\left(- i\phi' \left(\bar{u}^* \bar{u}_{\tau} - \bar{u}\bar{u}_{\tau}^*\right)-2v_R{\rm Re}(\overline{u})|\overline{u}|^2\right).
%\end{align}}
%\Revone{{\bf Remark}: Associated with the conservative part of the modified LL equation, we have the Hamiltonian density $\mathcal{H}=-\frac{1}{2}|\overline{u}|^4+|\overline{u}_{\tau}|^2+\left(\Delta+(\phi^{\prime})^2-2|v_s|^2\right)|\overline{u}|^2-i\phi' \left(\bar{u}^* \bar{u}_{\tau} - \bar{u}\bar{u}_{\tau}^* \right)-2v_R{\rm Re}(\overline{u})|\overline{u}|^2$}.
%
\\On the other hand, to construct the Lagrangian density leading to the non-conservative 
part of the LL equation, namely the right-hand side of Eq.~(\ref{LLNCVA}), we must 
find ${\mathcal{R}}$ such that
\begin{eqnarray}
\label{Folding-ncva}
{\left [ \frac{\partial {\overline{\mathcal{R}}}}{\partial \bar{u}_-^{*}} \right ]_{\rm PL}} ={-i\overline{u}+(-2|\overline{u}|^2+(\phi^{\prime})^2-{i}\phi^{\prime\prime})v_s-(v_s)^2\overline{u}^{*}-(\overline{u})^2v_s^{*}}
%& \Revone{-i\overline{u}-2iv_I|\overline{u}|^2+((\phi^{\prime})^2-{i}\phi^{\prime\prime})v_s-(v_s)^2\overline{u}^{*}+iv_I(\overline{u})^2},
\end{eqnarray}
%%,
and thus, an appropriate choice for the non-conservative terms yields
%
%% \Revone{}
\begin{eqnarray}
\label{Non-conservative-terms}
\overline{{\mathcal{R}}} =& \Big[ {-i\overline{u}_++(-2|\overline{u}_+|^2+(\phi^{\prime})^2-{i}\phi^{\prime\prime})v_s-(v_s)^2\overline{u}_+^{*}-(\overline{u}_+)^2v_s^{*}} \Big] {\bar{u}_-^{*}}+{c.c.},
\end{eqnarray}
where $c.c.$ denotes the complex conjugate of the preceding 
term [it is necessary to choose this term to be the 
integration constant that arises upon integrating 
Eq.~\eqref{Folding-ncva} to ensure the underlying NCVA 
parameters are real-valued].
Thus, the relevant Lagrangian density containing 
conservative and non-conservative terms can be written as 
\begin{eqnarray}
{\overline{\mathcal{L}}} = {\overline{{\mathcal{L}}}_{\rm cons}(\overline{u}_1)-\overline{{\mathcal{L}}}_{\rm cons}(\overline{u}_2)+\overline{\mathcal{R}}},
%\nonumber \\ 
%\nonumber
%&
 %+ \Big[ - i (1+\phi'') \bar{u}_+ + \left((\phi')^2 - 2 |\bar{u}_+|^2- \phi''\right) v_s 
%- (v_s)^2 \bar{u}_+^* - (\bar{u}_+)^2 v_s^* \Big] %\bar{u}_-,
\end{eqnarray}
where $\bar{u}_1 = (2\bar{u}_+ + \bar{u}_-)/2$ and $\bar{u}_2 = (2\bar{u}_+ - \bar{u}_-)/2$ and {the expression for $\overline{\mathcal{R}}$ is given in Eq.~(\ref{Non-conservative-terms}).
%
%From ${\mathcal{L}}$, we can derive, through the Euler-Lagrange 
%equations~(\ref{NCVAODE}) the full tweeze LL model at the PDE level. 
%
In order to obtain analytical insights into the dynamics of the model, our aim 
is to use an ansatz approximation of the intracavity field envelope {soliton} reducing 
its original Lagrangian to a Lagrangian over effective (yet slowly varying)
properties. Therefore, to approximate the CSs, we chose a six-parameter, 
$\vec{p}=(a,\xi,\sigma,b,c,d)$, Gaussian ansatz of the form:
\begin{eqnarray}
\bar{u}_j =& a_j \exp \left[ -\frac{(\tau - \xi_j)^2}{2\sigma_j^2} \right] 
\exp \left[ i (d_j (\tau - \xi_j)^2 + c_j (\tau - \xi_j) + b_j ) \right]
\label{6pAnsatzTweeze}
\end{eqnarray}
for $j=1$ and 2, where the CS variational parameters correspond to height $a$, 
center position $\xi$, width $\sigma$, phase $b$, velocity $c$, and chirp $d$. 
It is relevant to mention that, as shown in Ref.~\cite{JuliaSSB}, it is necessary
to add chirp into the ansatz since the original model is out of equilibrium.
As a consequence, for non-trivial solutions, there must exist
internal (fluid) flows from regions of effective gain to regions of effective loss. 
These balancing flows correspond, in turn, to spatial variations of the phase
that must be accounted by the chirp term.\footnote{The connection between phase
variations and (fluid) flows can be elucidated within setups corresponding
to the cGL equation with real coefficients; namely
NLS-type equations. Within NLS-type models, the Madelung transformation 
(that separates the wavefunction in density and phase terms) allows recasting 
the original NLS PDE into a modified (by the so-called quantum pressure term) 
non-viscous Eulerian fluid where, importantly, the gradient of the phase 
precisely corresponds to the fluid velocity~\cite{Dark_book}.}

Following the NCVA methodology for ansatz~(\ref{6pAnsatzTweeze}), we obtain a
system of ODEs [derived from Eq.~\eqref{NCVAODE}] for the variational parameters $\vec{p}$, which are listed in 
Appendix~\ref{AppendixA}. The resulting NCVA ODEs are cumbersome in their
explicit form in
that they include the terms $I_a$, $I_b$, $I_c$, $I_d$, $I_{\sigma}$, and $I_{\xi}$ 
which involve integrals. In order to simplify these integrals, we recast 
the ansatz of Eq.~(\ref{6pAnsatzTweeze}) using its amplitude and phase as follows
\begin{eqnarray}
\nonumber
{\bar{u}(\tau,\Vec{p}(z))} = \mathcal{A}(a, \sigma, \xi) \exp[i \Phi (b, c, d, \xi)],
\end{eqnarray}
where $\mathcal{A}(a,\sigma, \xi) \equiv a \exp (-{(\tau-\xi)^2}/{(2\sigma^2)})$ 
and $\Phi(b,c,d,\xi) \equiv b+c(\tau - \xi) + d (\tau-\xi)^2$ for variational 
parameters $\vec{p} = (a, b, c, d, \sigma, \xi)$. 
These integrals for all the variational parameters $p_i$ are of the form, {by leveraging the equivalence of the NCVA with the Kantorovitch method \cite{JuliaNCVA}:}
\begin{eqnarray}
\nonumber
{I_{p_i} = 2\, \mathrm{Re} \int_{-\infty}^{\infty}} {\Big[{-i\overline{u}+(-2|\overline{u}|^2+(\phi^{\prime})^2-{i}\phi^{\prime\prime})v_s-(v_s)^2\overline{u}^{*}-(\overline{u})^2v_s^{*}} \Big] \; \frac{\partial \bar{u}^{*}}{\partial p_i} \; d\tau.}
\end{eqnarray}
%
%\begin{eqnarray}
%\nonumber
%I_{p_i} = 2 \mathrm{Re} %\int_{-\infty}^{\infty} & %\Big[\left((\phi')^2 - 2 %|\bar{u}_{p_i}|^2- %\phi''\right) v_s 
%- (v_s)^2 \bar{u}_{p_i}^* - %(\bar{u}_{p_i})^2 v_s^* \Big] \; \frac{\partial \bar{u}^{*}}{\partial p_i} \; d\tau.
%\end{eqnarray}
%
Therefore, using the notation $\mathcal{A}_{p_i}= \partial \mathcal{A}/\partial {p_i}$,
$\Phi_{p_i}= \partial \Phi/ \partial p_i$, $v_I = \mathrm{Im}(v_s)$, 
$v_R = \mathrm{Re}(v_s)$, $\kappa = -4\mathcal{A}^2+2(\phi^{\prime})^2$ and  $\mathcal{X}^{(i)}=\frac{\partial \overline{u}^{*}}{\partial p_i}$,
together with ${\rm Re}(\mathcal{X}^{(i)})=\mathcal{X}_R^{(i)}$ and ${\rm Im}(\mathcal{X}^{(i)})=\mathcal{X}_I^{(i)}$, we can write 
a general form for $I_{p_i}$ as follows:
\begin{eqnarray}
\nonumber
{I_{p_i} =\int_{-\infty}^{\infty} \left(-2\mathcal{A}^2\Phi_{p_i}+\Gamma_1^{(i)}\cos(\Phi)+\Gamma_2^{(i)}\sin(\Phi)+\Gamma_3^{(i)}\cos(2\Phi)+\Gamma_4^{(i)}\sin(2\Phi)\right)d\tau},
\end{eqnarray}
where $\Gamma_{j}^{(i)}$, $j=1,2,3,4$ are defined through the following expressions
\begin{align}
    \Gamma_1^{(i)} =& \kappa v_R\mathcal{A}_{p_i}+\kappa v_I\mathcal{A}\Phi_{p_i}-2\phi^{\prime\prime}v_R\mathcal{A}\Phi_{p_i}+2\phi^{\prime\prime}v_I\mathcal{A}_{p_i}
    \\
    \nonumber 
    &+2(v_I^2-v_R^2)\mathcal{A}\mathcal{X}_R^{(i)}+4v_Iv_R\mathcal{A}\mathcal{X}_I^{(i)}
    \\[1.0ex]
     \Gamma_2^{(i)} =& -\kappa v_R\mathcal{A}\Phi_{p_i}+\kappa v_I \mathcal{A}_{p_i}-2\phi^{\prime\prime}v_R\mathcal{A}_{p_i}-2\phi^{\prime\prime}v_I \mathcal{A}\Phi_{p_i}
     \\
     \nonumber &+2(v_I^2-v_R^2)\mathcal{A}\mathcal{X}_I^{(i)}-4v_I v_R \mathcal{A}\mathcal{X}_R^{(i)}
     \\[1.0ex]
\Gamma_3^{(i)}=&-2\mathcal{A}^2(v_R\mathcal{X}_R^{(i)}+v_I\mathcal{X}_I^{(i)})
\\[1.0ex]
     \Gamma_4^{(i)}=& 2\mathcal{A}^2(v_R\mathcal{X}_I^{(i)}-v_I\mathcal{X}_R^{(i)})
\end{align}
%Furthermore, for the reference of the reader, expressions for $\mathcal{X}_R^{(i)}$ and $\mathcal{X}_I^{(i)}$ are provided below
%\begin{align}
 %   &\mathcal{X}_R^{(i)}=\mathcal{A}_{p_i}\cos(\Phi)-\mathcal{A}\Phi_{p_i}\sin(\Phi),\\
 %  &\mathcal{X}_I^{(i)}=-\mathcal{A}_{p_i}\sin(\Phi)-\mathcal{A}\Phi_{p_i}\cos(\Phi), 
%\end{align}
Succinctly, the expressions for $\mathcal{X}_R^{(i)}$ and $\mathcal{X}_I^{(i)}$ can be written as the following system:
\begin{align}
   \begin{bmatrix}
       \mathcal{X}_R^{(i)}\\\mathcal{X}_I^{(i)}
   \end{bmatrix} =\begin{bmatrix}
       \cos(-\Phi)&&-\sin(-\Phi)\\\sin(-\Phi)&&\cos(-\Phi)
   \end{bmatrix}\begin{bmatrix}
       \mathcal{A}_{p_{i}}\\-\mathcal{A}\Phi_{p_{i}}
   \end{bmatrix}.
\end{align}
%
%\begin{eqnarray}}
%I_{p_i} &=\int_{-\infty}^{\infty} \; d\tau \Big( %\left(\mathcal{X} - 3\mathcal{A}^2\right) \mathcal{A}_{p_i} %\left[ 2 v_r \cos(\Phi) + 2 v_i \sin(\Phi) \right] \nonumber \\
%&- \mathcal{A}\mathcal{A}_{p_i} \left[2v_r^2 \cos(2\Phi) - 2 %v_i^2\cos(2\Phi) + 4 v_i v_r \sin(2\Phi) \right] \nonumber \\
%&+ \left( \mathcal{X} - \mathcal{A}^2\right) \mathcal{A} %\Phi_{p_i} \left[ 2 v_r \cos(\Phi) - 2 v_i \sin(\Phi) \right] %\nonumber \\
%\nonumber
%&- \mathcal{A}^2\Phi_{p_i} \left[2v_i^2 \sin(2\Phi) - 2 v_r^2\sin(2\Phi) + 4 v_i v_r \cos(2\Phi) \right] 
%\Big).
%\end{eqnarray}
%
Although $I_{p_i}$ may appear cumbersome, the integrals are reduced by the presence
(or not) of the derivatives $\mathcal{A}_{p_i}$ and $\Phi_{p_i}$, 
e.g., $\mathcal{A}_b = \mathcal{A}_c = \mathcal{A}_d = 0$ and 
$\Phi_a = \Phi_{\sigma} = 0$. The only derivatives that are of importance are:
\begin{eqnarray}
&\mathcal{A}_a = \displaystyle\frac{\mathcal{A}}{a}, 
\quad
\mathcal{A}_{\sigma} = \displaystyle\frac{(\tau - \xi)^2}{\sigma^3} \mathcal{A}, 
\quad
\mathcal{A}_{\xi} = \displaystyle\frac{\tau-\xi}{\sigma^2} \mathcal{A}, 
\nonumber 
\\[1.0ex]
\nonumber 
&\Phi_b = 1,
\quad
\Phi_c = \tau-\xi, 
\quad
\Phi_{\xi} = -c - 2d(\tau - \xi), 
\quad
\Phi_d = (\tau- \xi)^2. 
\end{eqnarray}
All the relevant integrals are approximated numerically 
to spectral accuracy with the trapezoidal rule (see 
Ref.~\cite{QUAR}).

Let us now use the NCVA equations of motion both to seek to capture the
stationary CS states, as well as to describe the dynamics of CSs
subject to the tweezing in Eq.~(\ref{tau0}) in order to assess its tweezability
as a function of $\tau_f$ (how far we want to tweeze) and $\beta$ (how adiabatic
is the tweezing). 

%%%%%%%%%%%%%%%%%%%%%%%%%%%%%%%%%%%%%%%%%%%%%%%%%%%%%%%
\begin{figure}
    \centering
    \includegraphics[width=0.5\linewidth]{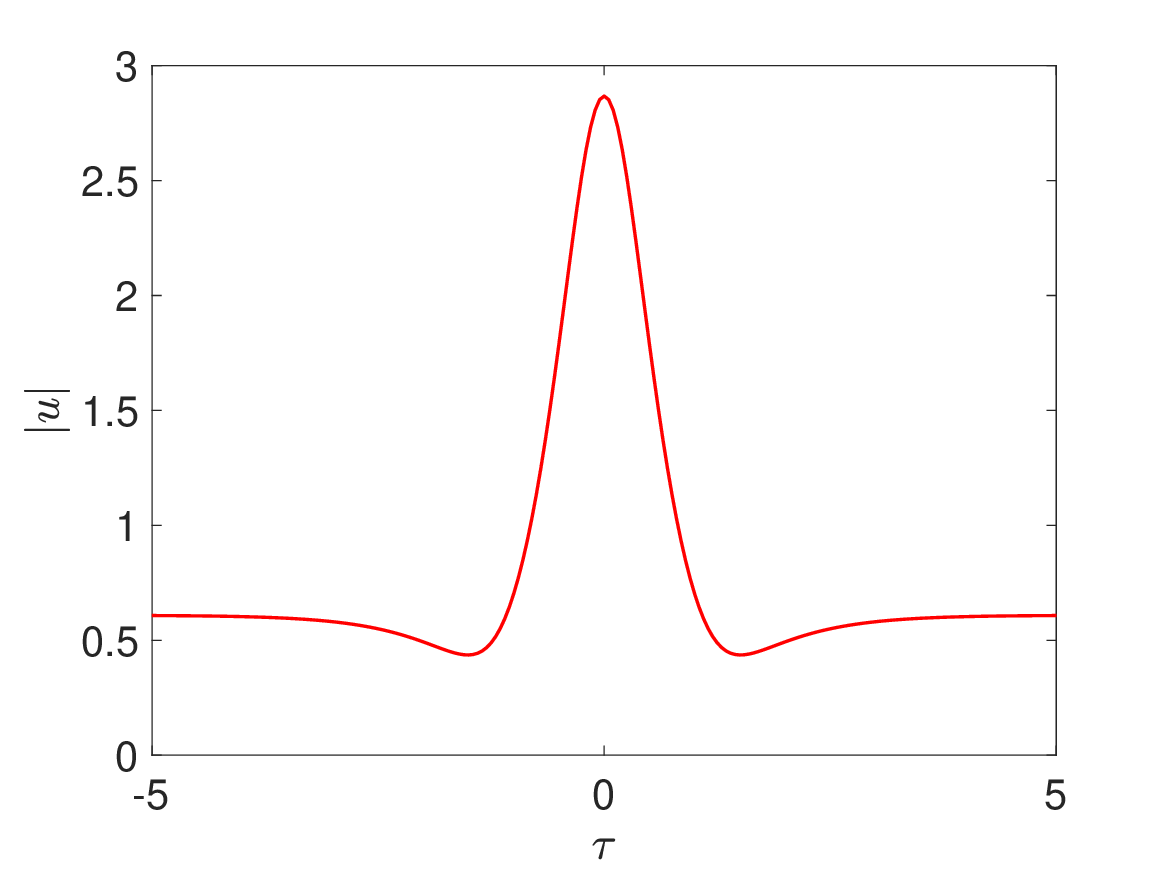}%{BULK_FULL_PROF.eps}
    \caption{A LL CS profile (where the profile of $|u|$ is shown) recovered for parameters $\Delta=3.5$ and $u_{\rm in}=2$ in the absence of the phase modulation $\phi(\tau) \equiv 0$.}
    \label{fig:2}
\end{figure}
%%%%%%%%%%%%%%%%%%%%%%%%%%%%%%%%%%%%%%%%%%%%%%%%%%%%%%%

%%%%%%%%%%%%%%%%%%%%%%%%%%%%%%%%%%%%%%%%%%%%%%%%%%%%%%%%%%%%%%%%%
\section{Results}
\label{section:TweezeResults}
%%%%%%%%%%%%%%%%%%%%%%%%%%%%%%%%%%%%%%%%%%%%%%%%%%%%%%%%%%%%%%%%%

%%%%%%%%%%%%%%%%%%%%%%%%%%%%%%%%%%%%%%%%%%%%%%%%%%%%%%%%%%%%%%%%%%
\subsection{Determination of (non-)tweezed stable CS states}
%%%%%%%%%%%%%%%%%%%%%%%%%%%%%%%%%%%%%%%%%%%%%%%%%%%%%%%%%%%%%%%%%%

As a precursor to performing the temporal tweezing study, we first identify an appropriate regime in the parameters (i) ${u}_{\rm in}$ and $\Delta$ and subsequently the phase modulation parameters 
(ii) $h_\phi$ and $\sigma_\phi$ [in Eq.~\eqref{phi}].
By the term ``appropriate'' here, we mean a regime
 where the CS state is spectrally stable both in the absence
 of the tweezer (in the bulk LL setting), as well as in the presence
 of the tweezer, so that it be amenable to dynamical tweezing.
 Crucially, this determination leads to suitable stable CS states which can be successfully temporally tweezed
 for some non-trivial regime in the adiabaticity
 ($\beta$) and final fast time ($\tau_f$) parameters
[defined by the temporal tweezing of Eq.~\eqref{tau0}]. 

\noindent
Two remarks are in order:
\begin{itemize}
\item[(a)]
As alluded to earlier, the computation of fixed points to the ODE of Eq.~\eqref{LLNoSol} is initiated by first fixing the homogeneous background pedestal $v_s$ [obtained as a solution to Eq.~\eqref{steadyStateLL}] with a choice of the parameters $\Delta$ and $u_{\rm in}$. Notably, this background state $v_s$ stays fixed in the presence of the localized phase modulation $\phi(\tau)$ as well [cf.~Eq.~\eqref{LLConstraint}]. Thereafter, we employ the Newton-conjugate gradient iterative scheme \cite{JNCG} to recover the CS profiles in $\bar u_0(\tau)= u_0(\tau)-v_s$. To study the stability of the CS, we  linearize Eq.~\eqref{LLNCVA} around these stationary states by inserting the ansatz $\bar u= \bar u_0+\epsilon( f(\tau)\exp(i\omega z)+g^{*}(\tau)\exp(-i\omega^{*}z))$, for $\epsilon\ll 1$, which yields the following linear stability (eigenfrequency) problem (at order $\epsilon$) upon collecting the coefficients of $\exp(i\omega z)$ and $\exp(-i\omega^{*} z)$
    \begin{equation}
        \label{Lin-stab-op}\begin{bmatrix}
A+\partial_{\tau\tau}&&B
\\[1.0ex]
-B^{*}&&-A^{*}-\partial_{\tau\tau}
\end{bmatrix}\begin{bmatrix}f
\\[1,0ex]
g\end{bmatrix}=\omega \begin{bmatrix}f
\\[1.0ex]
g\end{bmatrix},
    \end{equation}
    where $A\equiv 2|\bar u_0|^2-(\Delta+(\phi^{\prime})^2-2|v_s|^2)+i\phi^{\prime\prime}+i+2v_s\bar u_0^{*}+2v_s^{*}\bar u_0$ and $B\equiv \bar u_0^2+2v_s\bar u_0+v_s^2$, for the eigenfrequency
    $\omega$.
 \medskip   
    
\item[(b)]
    We complement our studies at the level of the LL Eq.~\eqref{eq:LLETweeze} with analogous investigations at the level of the NCVA ODE system (see Appendix~\ref{AppendixA}). We compute fixed points to this ODE system $\dot {\vec{p}}=\vec{F}(\vec{p})$ by solving a six-dimensional algebraic system $F(\vec{p})=0$ using a Newton-based black box-solver {\tt nsoli.m} \cite{CTK}. The NCVA eigenfrequencies $\omega_0$
    %$\omega_0=-i\lambda$ 
    of the corresponding $6\!\times\! 6$ Jacobian 
    ${\partial{\vec F}}/{\partial{\vec p}}$ can also be used as an appropriate diagnostic to characterize linear stability of stationary states, and thus predict successful temporal tweezing. Importantly, also, the resulting 
    values of $\omega_0$ are directly compared with the spectral stability findings
    at the level of the LL PDE, as yet another layer of comparison regarding
    the accuracy of the NCVA approach.
\end{itemize}

%%%%%%%%%%%%%%%%%%%%%%%%%%%%%%%%%%%%%%%%%%%%%%%%%%%%%%%
\begin{figure}
    \centering
    \includegraphics[width=0.99\linewidth]{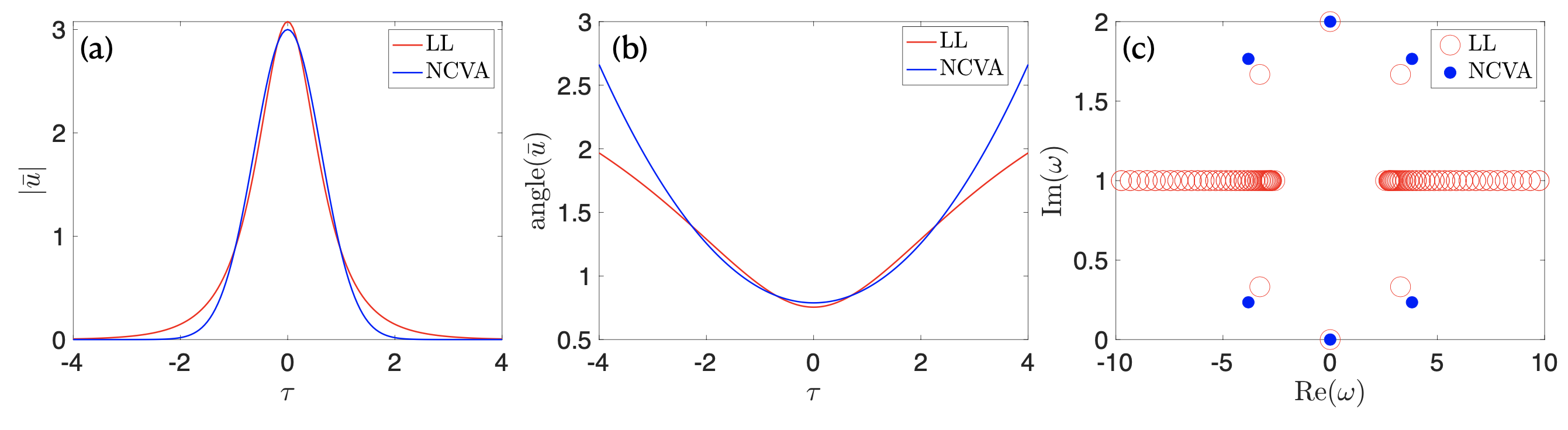}%{Bulk_LL_NCVA.png}
    \caption{A comparison of the CS profile at the NCVA and LL PDE levels for parameters $\Delta=3.5$ and $u_{\rm in}=2$ in the absence of the phase modulation $\phi \equiv 0$. 
    Comparison of the (a) amplitude and (b) phase 
    of the CS for the LL model (red) and its
    NCVA counterpart (blue).
    (c) Stability spectra of the CS for the LL model
    (red circles) together with its NCVA counterpart 
    (blue dots).
    }
    \label{fig:3}
\end{figure}
%%%%%%%%%%%%%%%%%%%%%%%%%%%%%%%%%%%%%%%%%%%%%%%%%%%%%%%

After an extensive line search in $\Delta$ (for fixed $u_{\rm in}=2$), we find that for $\Delta \gg \sqrt{3}$, stable CS are recovered for $\phi(\tau)\equiv 0$. Furthermore, their stability is reflected at the level of the NCVA ODE system as well.
The profile ($|u|$) of such a LL CS is depicted in Fig.~\ref{fig:2}. 
Next, we showcase the ability of the NCVA approach to capture the statics and stability of the original CSs in the LL equation.
In particular, panels (a) and (b) if Fig.~\ref{fig:3}
depict a typical comparison between shape and phrase profiles of the NCVA and LL CS.
The figure evidenced that the NCVA is able
to give a good approximation of the exact CSs.
Furthermore, panel (c) depicts the associated linear stability spectra $\omega$ for both the NCVA and the LL CS.
%F
Note that for the choice of parameters, the spectra lie entirely above the lower half plane,
i.e., have ${\rm Im}(\omega)\geq 0$, confirming stability. 
Additionally, and
reassuringly, the stability eigenvalues of the NCVA ODE system (blue dots) are also seen 
to demonstrate reasonable quantitative agreement with the discrete spectra of the original LL CS.

%%%%%%%%%%%%%%%%%%%%%%%%%%%%%%%%%%%%%%%%%%%%%%%%%%%%%%%
\begin{figure}
    \centering
    \includegraphics[width=0.5\linewidth]{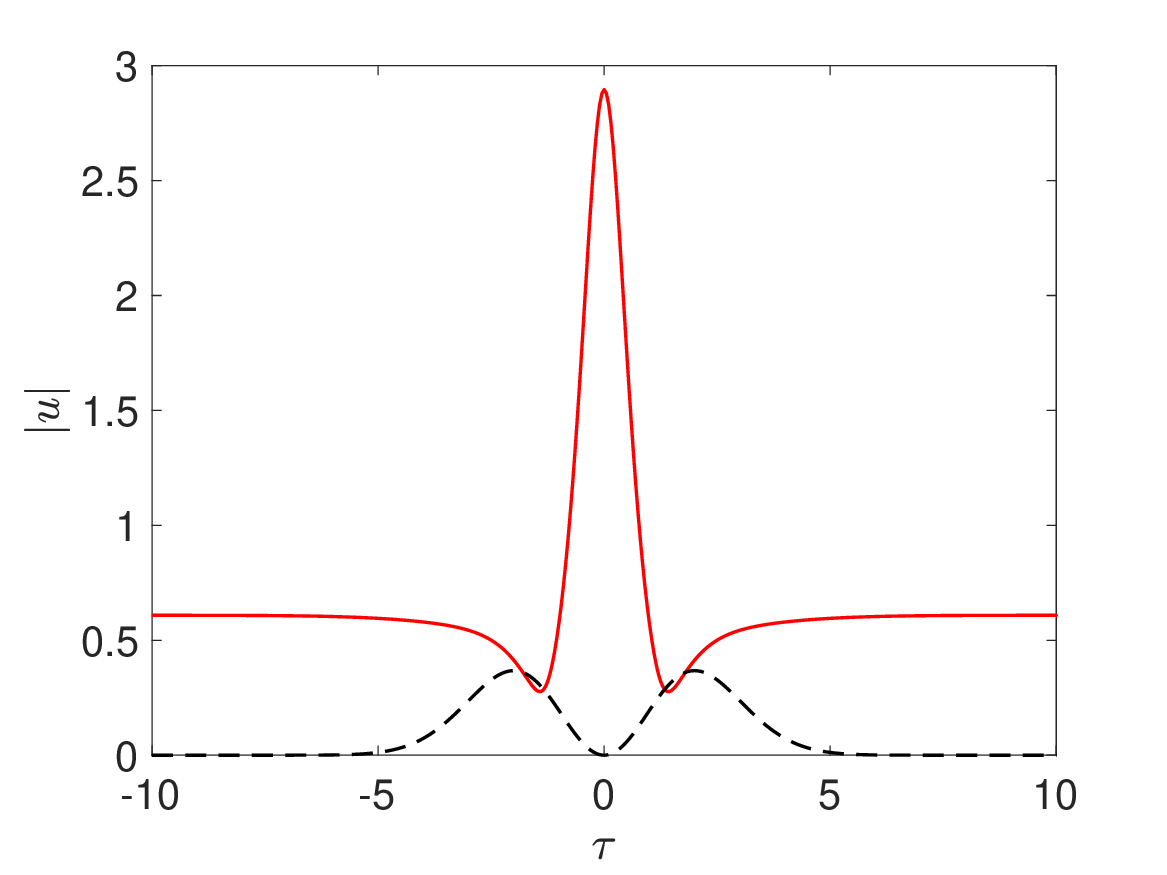}%{Tweezed_CS_full_profile.eps}
    \caption{A stationary LL CS profile (in terms of $|u|$; see red curve) 
    in the presence of the effective potential (see black dashed curve) 
    ensuing from the phase modulation spatial profile. 
    This case corresponds to the parameters $\Delta=3.5$, $u_{\rm in}=2$,
    $h_\phi=2$, and $\sigma_\phi=2$ in the presence of the phase 
    modulation $\phi(\tau)$ described in Eq.~\eqref{phi}.}
    \label{fig:4}
\end{figure}
%%%%%%%%%%%%%%%%%%%%%%%%%%%%%%%%%%%%%%%%%%%%%%%%%%%%%%%

%%%%%%%%%%%%%%%%%%%%%%%%%%%%%%%%%%%%%%%%%%%%%%%%%%%%%%%
\begin{figure}
    \centering
    \includegraphics[width=0.99\linewidth]{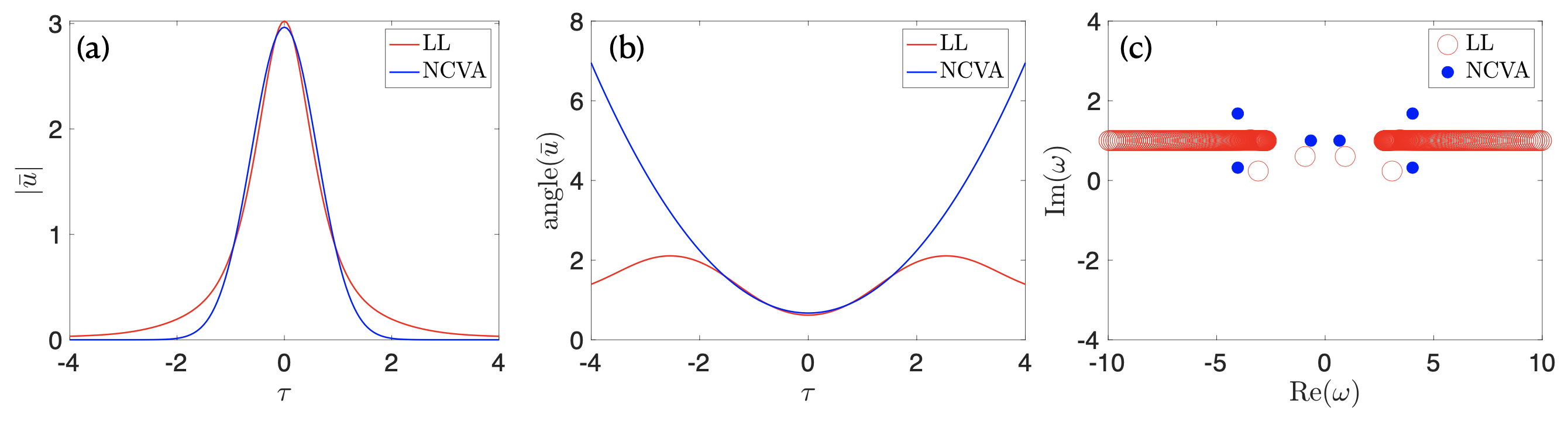}%{Tweezed_CS.png}
    \caption{A comparison of the stationary CS profile (with the 
    pedestal factored out) in the
    presence of the phase modulation, at the NCVA and LL PDE levels for parameters $\Delta=3.5$, $u_{\rm in}=2$, $h_{\phi}=\sigma_{\phi}=2$: Comparisons  (a)  of the amplitude variations of the LL CS (in red), and the NCVA counterpart in blue, (b) the phase variations of the LL CS (in red) and the NCVA counterpart in blue and (c) the LL stability spectra plotted against the NCVA stability eigenvalues.}
    \label{fig:5}
\end{figure}
%%%%%%%%%%%%%%%%%%%%%%%%%%%%%%%%%%%%%%%%%%%%%%%%%%%%%%%

Next, we consider the case of adiabatically turning on 
the phase modulation, in the presence of a bulk 
($\phi\equiv 0$) LL CS having fixed $\Delta=3.5$
and ${ u}_{\rm in}=2$. 
Figure~\ref{fig:4} depicts a typical tweezed CS and the
corresponding effective trapping potential originating from
the phase modulation.
We find that for shallow 
effective potential, i.e. $\underset{\tau}{\rm max}(\phi^{\prime})^2\ll 1$, 
both the NCVA and the LL models
support CSs that are stable as showcased 
in Fig.~\ref{fig:5}. This stability of 
the confined CS (within the phase modulation) is crucial towards a potential successful tweezing of such
localized states.
As for the untrapped case, the NCVA predicts 
stability for this trapped CSs.
For the remainder of the paper, we fix $h_\phi=2$ and $\sigma_{\phi}=2$. 
The choice of these phase modulation parameters will allow us to 
demonstrate successful not only
static, but also dynamic tweezability, as we will now show.

%%%%%%%%%%%%%%%%%%%%%%%%%%%%%%%%%%%%%%%%%%%%%%%%%%%%%%%
\begin{figure}[t!]
\centering
%\centerline
\includegraphics[width=0.45\textwidth]{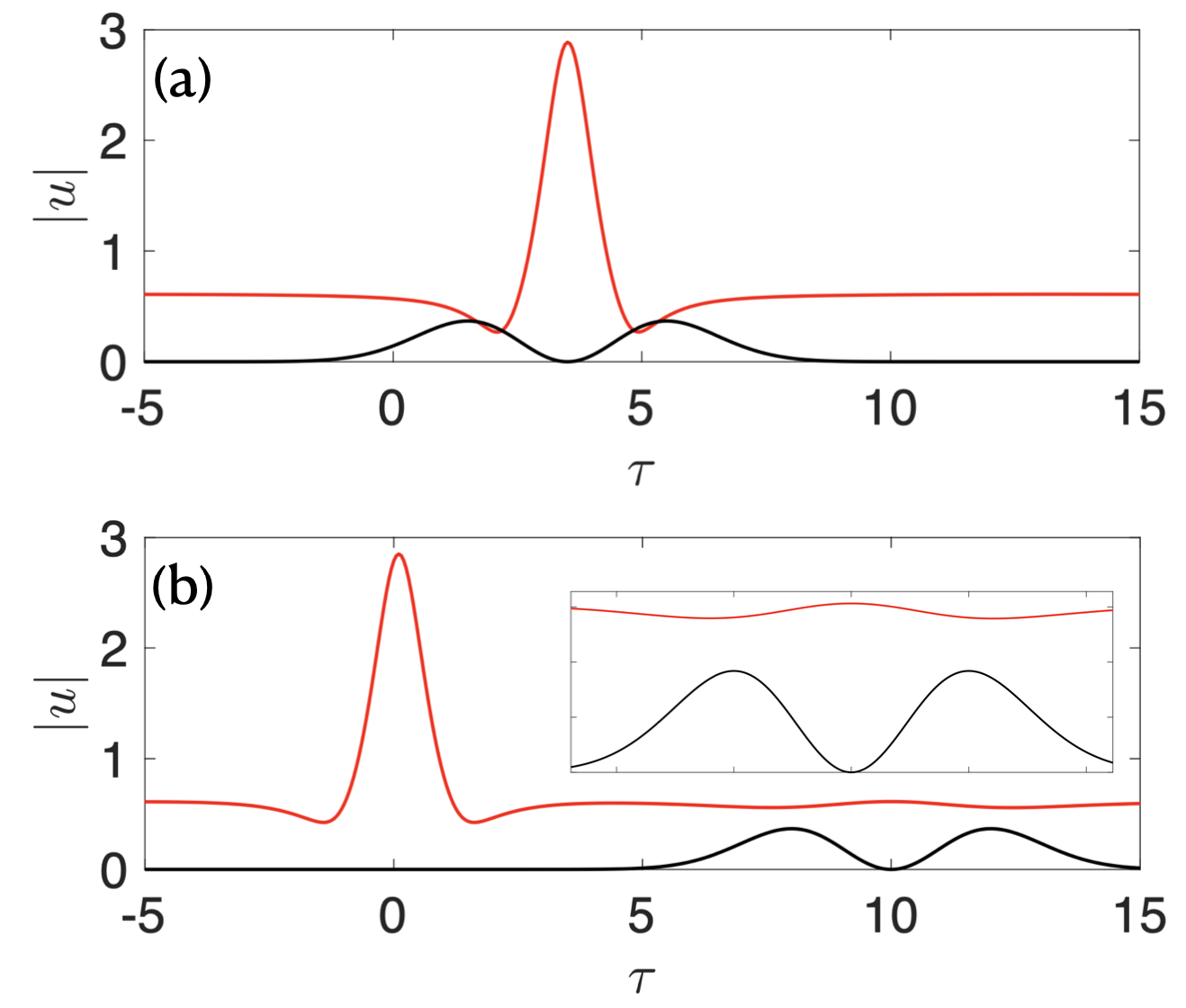}%{tweezing_scenarios_final.png}
%{Tweezing_scanarios.png}
% \rule{35em}{0.5pt}
\caption[Temporal Profiles of Fundamental States]{Temporal profiles of the 
intracavity field envelope $|u|^2$ (red line) for the two fundamental 
CS states of the system. 
In this example, the effective potential, 
$V_{\rm eff} = (\phi')^2$ (solid black line) is varying in the
evolution variable
in the {two} panels with parameters $\sigma_\phi = 2$ and $h_\phi = 2$. 
The top panel is an example of a trapped CS by the time-dependent 
effective potential (i.e., a successfully tweezed CS). The bottom panel 
is an example of a failed tweezing attempt. {Note that in this case, 
a small amplitude localized state is observed within the tweezing 
potential, which can be viewed as a modulation of the homogeneous 
background pedestal.}
}
\label{fig:threeStates}
\end{figure}
%%%%%%%%%%%%%%%%%%%%%%%%%%%%%%%%%%%%%%%%%%%%%%%%%%%%%%%

%%%%%%%%%%%%%%%%%%%%%%%%%%%%%%%%%%%%%%%%%%%%%%%%%%%%%%%
\begin{figure}
    \centering
    \includegraphics[width=0.99\linewidth]{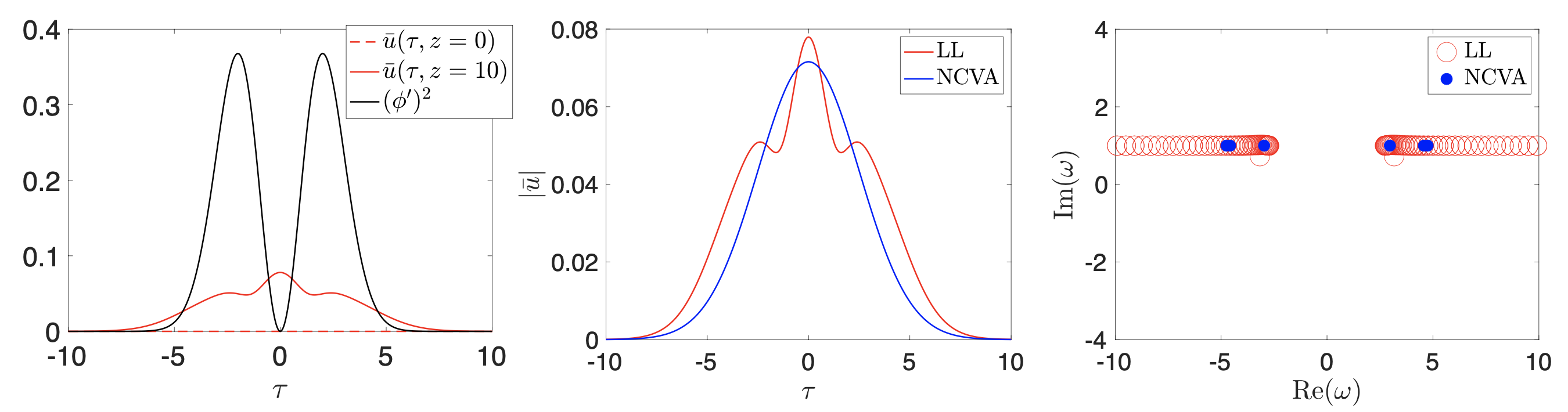}%{MH_CS.png}
    \caption{(a) A localized state arises within the effective potential (shown in black solid line) as a steady state past $z=10$, (red solid line) when initialized with the homogeneous pedestal (red dashed line). Here the field variable $|\bar u|=|u-v_s|$ is shown to describe the waveforms. (b) The NCVA successfully captures this localized state: the panel shows the amplitude profile of the localized modulation at the level of the LL (red solid) and its NCVA analog (blue solid).
    (c) The LL and NCVA stability spectra (eigenvalues) confirm the 
    stability of this localized feature within
    the confining potential. The absence of point spectrum modes here indicates
    that this is the analog of the homogeneous state in the presence of the
    phase-modulation-induced confinement.}
    \label{fig:smallamp}
\end{figure}
%%%%%%%%%%%%%%%%%%%%%%%%%%%%%%%%%%%%%%%%%%%%%%%%%%%%%%%

%%%%%%%%%%%%%%%%%%%%%%%%%%%%%%%%%%%%%%%%
\subsection{Temporal tweezing}
%%%%%%%%%%%%%%%%%%%%%%%%%%%%%%%%%%%%%%%%

We now explore the existence and dynamical properties for a tweezed cavity soliton in
the effective potential described by $V_{\rm eff} = (\phi')^2$ (also referred
to as the ``tweezer''). 
Before embarking on the dynamical tweezability of CSs, we must know what are the 
different possible scenarios. In particular, for a tweezer there are two fundamental states corresponding to the following 
qualitatively different scenarios:
(a) a CS inside the tweezer
(also of relevance to the case of successful dynamical tweezing) and
(b) a bulk CS outside of the tweezer, arising also in the setting 
of non-tweezed (or unsuccessfully tweezed) CS.
These two scenarios are depicted in Fig.~\ref{fig:threeStates} for 
$\sigma_\phi = 2$ and $h_\phi=2$ (which corresponds to a shallow tweezer). 
Interestingly, in the absence of the CS within the tweezer, a very small 
amplitude, localized, modulation of the background emerges within the
effective potential created by the phase modulation 
[see Fig.~\ref{fig:smallamp}(a)]. Notably, this small amplitude 
modulation of the background can also be captured by the Gaussian 
NCVA ansatz, as shown in Fig.~\ref{fig:smallamp}(b), and is a 
stable waveform, as illustrated in Fig.~\ref{fig:smallamp}(c).

If we now consider a dynamically varying tweezer [i.e., $\tau_0$ given by Eq.~(\ref{tau0})],
for constant values of $\sigma_\phi$ and $h_\phi$, there will be critical values of 
$\beta$ and $\tau_f$ in Eq.~(\ref{tau0}) corresponding to thresholds between the non-tweezed and the tweezed
fundamental states. 
We now proceed to provide a characterization of the regions of 
existence of these two states for the full LL Eq.~(\ref{eq:LLETweeze}) 
and compare to the results of the NCVA 
with respect to the parameters $\beta$ and $\tau_f$. 
In what follows, we study 
different tweezing possibilities over the parameter space spanned by $\tau_f$, 
the final temporal displacement, and $\beta$, the degree of adiabaticity for 
the speed of the tweezer. A successfully tweezed CS will correspond to a CS 
that stays inside the effective potential as the tweezer is displaced, while
a failed attempt will result in the CS being ``left behind'' the dynamically
evolving phase modulation profile (and, in particular, its effective 
confinement region).

Rather than analyzing the individual evolution of temporal profiles, we can 
express the various states by the power contained inside and outside of the 
effective confinement region of the phase modulation profile. For the full LL model, the CS temporal density $\rho = |u|^2$ 
and the homogeneous state density $\rho_0 = |v_s|^2$ can express the power inside the tweezer $P_{\rm I}(z)$ and outside the tweezer $P_{\rm O}(z)$, 
which are given respectively by 
\begin{subequations}
\label{PInPOut}
\begin{eqnarray}
P_{\rm I}(z) = \int_{\mathcal D} (\rho - \rho_0) d\tau, \label{Pin} \\[1.0ex]
P_{\rm O}(z) = \int_{\bar{\mathcal D}} (\rho - \rho_0) d\tau, \label{Pout}
\end{eqnarray}
\end{subequations}
where {${\mathcal D}$
%$ = [ -2\sigma, 2\sigma]$} \RED{(I think it is more correct for this to be twice the tweezer width. This is because the tweezed CS is manipulated with partial success even when a reasonable portion of the soliton mass falls outside the tweezer width. Incidentally, Julia's codes implicitly uses this criteria as well.)} 
is the domain of the tweezer which is 
defined as a symmetric interval around $\tau_0$ extending to twice the tweezer 
width and $\bar{\mathcal D}$ is the complement of ${\mathcal D}$. Note that we
subtracted the background pedestal $\rho_0=|v_s|^2$, in order to 
remove the effects of the 
pedestal (constant steady state background) and ensure the power for a CS is positive. 
We thus define the total power of the solution by
\begin{eqnarray}
\nonumber
P_{\rm Tot} = \int_{-\infty}^{+\infty} (\rho - \rho_0) d\tau = P_{\rm I}+P_{\rm O}.
\end{eqnarray}
For the NCVA, we use the variational parameters 
and construct the CS temporal density $\bar{\rho} = |\bar{u}|^2$ and extract the 
power inside the tweezer $P_{\rm I}(z)$ and outside the tweezer $P_{\rm O}(z)$ 
which are given, respectively, by 
\begin{eqnarray}
\bar{P}_{\rm I}(z) = \int_{\mathcal D} \bar{\rho }d\tau, 
\nonumber
\\[1.0ex]
\nonumber
\bar{P}_{\rm O}(z) = \int_{\bar{\mathcal D}} \bar{\rho } d\tau, 
\end{eqnarray} 
where ${\mathcal D}$ and $\bar{{\mathcal D}}$ are the same intervals described above
and recall that in the construction of the NCVA ansatz we already subtracted out the 
effects of the background pedestal. 
The relative power change during the tweezing, namely between $z=0$ and $z=z_f$,
is then defined by
\begin{subequations}
\label{QInQOut}
\begin{eqnarray}
Q_{\rm I} = \frac{P_{\rm I}(0) - P_{\rm I}(z_f)}{P_{\rm Tot} },\label{QIn} \\[1.0ex]
Q_{\rm O} = \frac{P_{\rm O}(0) - P_{\rm O}(z_f)}{P_{\rm Tot} },
\label{QOut}
\end{eqnarray}
\end{subequations}
for, respectively, the inside and outside of the tweezer. The same definition is
used for the relative power changes ensuing from the NCVA approach.

By considering $Q_{\rm I}$ and $Q_{\rm O}$, we can effectively find the thresholds 
between the various states in the relevant $(\beta,\tau_f)$ parameter space. 
For example, if we begin with a steady state CS inside the tweezer at 
$\tau_0=0$ and move with a given $\beta$ and $\tau_f$, then a successfully 
tweezed CS will result on approximately the same powers at $z=0$ as $z = z_f$ 
such that $P_{\rm I}(0)\approx P_{\rm I}(z_f)$ as well as 
$P_{\rm O}(0) \approx P_{\rm O}(z_f)$, and thus resulting in power 
ratios $Q_{\rm I} \approx 0 \approx Q_{\rm O}$. 
On the contrary, if at the end of the tweezing, a bulk CS state is left in the wake of the tweezer, the resulting 
power changes will yield $Q_{\rm I} \approx 1$ and $Q_{\rm O}\approx -1$.

For our analysis, as mentioned in the previous subsection, stability 
considerations limit us to choosing shallow tweezing potentials, namely 
effective potentials with small to moderate heights characterized by 
$h_{\phi}$. For this work, we fix $h_\phi=2$.
In future efforts, we will provide a more systematic analysis of the
stability considerations for different phase modulation profile parameters
and explore the possibilities of static and dynamic tweezing accordingly.
Here, we restrict our considerations to a proof-of-principle example
of a phase modulation enabling a stable static confined CS and examine the 
dynamical outcome for different parameters of the selected protocol
$\tau_0(z)$.
Moreover, we reiterate that all examples we discuss are limited to 
fixing $u_{\rm in} = 2$ and $\Delta =3.5$.

We now report on the results having fixed the tweezer width to 
$\sigma_\phi = 2$, as mentioned earlier. 
The parameter space for $\beta$ and $\tau_f$ is discretized into 41 points 
between 0.1 and 20 in both directions, giving 1681 combinations for 
$\tau_0(z)$ as per Eq.~(\ref{tau0}). 
The full LL Eq.~(\ref{eq:LLETweeze}) is solved using a standard second order finite 
difference scheme in space and a standard fourth order Runge-Kutta method in time.
It is relevant to note that the integration of the NCVA system of ODEs required 
the use of a stiff ODE solver and thus we used Matlab's {\tt ode15s}.
Both the full LL PDE and the NCVA ODEs are evolved until $z_f = 4z^*$, 
with $z^* = 2.5$ (see Fig.~\ref{fig:tau0}) which ensures that the successfully 
tweezed or non-tweezed CS had enough time to converge towards 
their respective steady states.

%%%%%%%%%%%%%%%%%%%%%%%%%%%%%%%%%%%%%%%%%%%%%%%%%%%%%%%
\begin{figure}
    \centering
\includegraphics[width=0.8\linewidth]{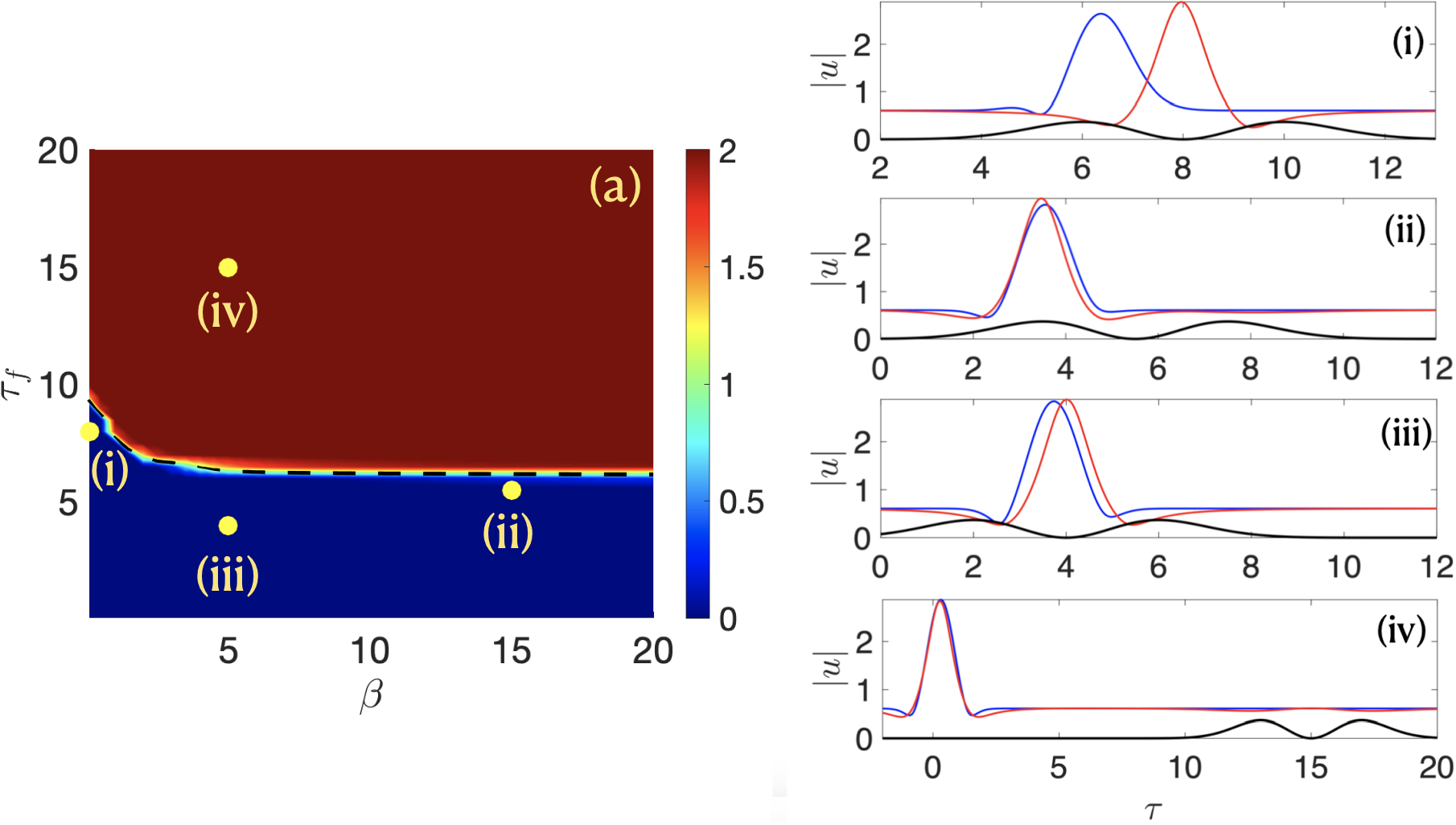}
%{tweezed_map_CS_2.png}%{tweezing_map.png}    
\caption{(a) The temporal tweezing map (surface plot for the tweezability 
index $\Delta Q$), generated from the LL PDE calculations for
$z^{*}=2.5$. On this map we overlay the contour $\Delta Q=1$ 
(see black dashed line) extracted from the analogous map generated 
from the NCVA computations, demonstrating excellent agreement. 
The right panels correspond to four representative cases depicting the
amplitude profiles of the NCVA (blue solid line) and LL PDE (red solid line) 
at $z=10$ alongside the effective potential profile (black solid line).
The values $(\beta,\tau_f)$ values for each case are depicted by the
yellow points.
The four representative cases are as follows.
Panels (i) and (ii) depict two marginally tweezed CS for 
$(\beta,\tau_f)=(0.1,8)$  and $(\beta,\tau_f)=(15,5.5)$ (see text for details).
Panel (iii) corresponds the case $(\beta,\tau_f)=(5,4)$ well within 
the tweezability region.
Panel (iv) corresponds to $(\beta,\tau_f)=(5,15)$ well within the
(red) non-tweezed region.}    
\label{fig:tweezing-map}
\end{figure}
%%%%%%%%%%%%%%%%%%%%%%%%%%%%%%%%%%%%%%%%%%%%%%%%%%%%%%%

We will now focus on following the power ratios $Q_{\rm I}$ and $Q_{\rm O}$ from 
Eqs.~(\ref{QInQOut}) for the different parameter combinations.  
In order to identify the different dynamical regions, 
both $Q_{\rm I}$ and $Q_{\rm O}$ need to be analyzed simultaneously. Therefore, for a more compact
interpretation of the results, we use the difference in power ratios, referred to as the tweezability index
$\Delta Q = Q_{\rm I} - Q_{\rm O}$, such that its values represent
the following dynamical tweezing scenarios:
\begin{itemize}
\item[(i)]
$~\Delta Q \approx 0$: successful tweezing,
\item[(ii)]
$~\Delta Q \approx 2$: failed tweezing, whereby 
the CS was left behind by the tweezer.
\end{itemize}
Figure~\ref{fig:tweezing-map}(a) depicts $\Delta Q$ as obtained from 
the LL PDE calculations as a function of $\beta$ and $\tau_f$ 
where the above two different tweezing regions are clearly defined:
(i) successful tweezing ($\Delta Q \approx 0$) in blue and
(ii) failed tweezing of the CS ($\Delta Q \approx 2$) in red. 
We note the sharp transition between these two fundamental 
states that could result from the dynamical tweezing. 
In addition to the tweezing map for the LL CS, panel (a) also depicts
the overlaid contour $\Delta Q=1$ from the NCVA reduction (see dashed
black line).
We note the excellent agreement [up to the chosen discretization in 
the $(\beta,\tau_f)$ plane] between the full LL tweezing map and
the prediction of the reduced NCVA approach.
Interestingly the boundary transition between tweezed and non-tweezed 
states appears to rapidly asymptote to $\tau_f\approx 5.5$ as
the degree of adiabaticity $\beta$ is increased.
The behavior of $\Delta Q$ depicted in for Fig.~\ref{fig:tweezing-map}(a) 
confirms the existence of two fundamental scenarios (at both LL and NCVA 
levels): a tweezed CS for all $\beta$ when $\tau_f \lessapprox 5.5$ 
(blue region) and a non-tweezed CS (red regions).
The right subpanels in Fig.~\ref{fig:tweezing-map} depict four different
representative cases after the temporal tweezing attempt where the
blue and red lines represent, respectively, the LL and NCVA CS while
the effective potential is depicted in black.
Panel (i) corresponds to the point in the left corner of the tweezed 
region $(\beta,\tau_f)=(0.1,8)$ in the tweezing map. In this extreme
example, although the the CS is successfully tweezed (i.e. confined 
within the region defined by the effective potential), the steering 
efficiency, i.e. alignment between $\tau_0(z)$ and $\xi(z)$ is not as 
effective. The success of the tweezing in turn is at the level of 
the integrated quantity $\Delta Q$, while we note the lack of 
perfect agreement in terms of soliton positions as predicted by 
the variational formulation and the LL PDE computation. 
Furthermore, we observe a maximal quantitative differences between 
the soliton centers as predicted by the NCVA and LL in this case. 
Panel (ii) corresponds to a CS that is tweezed with a large degree of 
adiabaticity $\beta=15$ and for $\tau_f=5.5$. The amplitude of the 
waveforms for the LL and the NCVA CS at $z=10$ demonstrate very good 
agreement in the profiles and their corresponding soliton 
centers [$\xi(z=10)\approx 3.7$]. For this transitional case, 
we note that upon evolution to a longer fast time ($z$), the LL 
and NCVA CS eventually catch up to the effective potential center. 
Panel (iii) corresponds to tweezing parameters $(\beta,\tau_f)=(5,4)$ 
well within the (blue) tweezability region. In this case, there is 
reasonable agreement between the respective LL and NCVA profiles.
Panel (iv) corresponds corresponds to the choice $(\beta=,\tau_f)=(5,15)$
well inside the (red) non-tweezed. Here both LL and NCVA CS waveforms 
at $z=10$ are left beghind in the wake of the tweezer, with, again,
very good mutual agreement in the soliton position $\xi(z)$.

The results presented above evidence that, with the current parameter 
values that we used, a combination of $0<\beta<20$ with $\tau_f \lesssim 4$
ensures a proper, successful, and efficient [in terms of alignment of 
$\xi(z)$ and $\tau_0(z)$] tweezing of the CS. 
Furthermore, we note that the NCVA provides an excellent representation of 
the full PDE dynamics. This is particularly of interest, as the NCVA 
calculations (ODE system) are far less costly than the full PDE computations,
corresponding to a low (six-) dimensional representation thereof.

%%%%%%%%%% Fig %%%%%%%%%%%%%%%%%%%%%%%%%%%%%%%%%%%%%%
\begin{figure}[t!]
\centering
\includegraphics[width=0.8\textwidth]{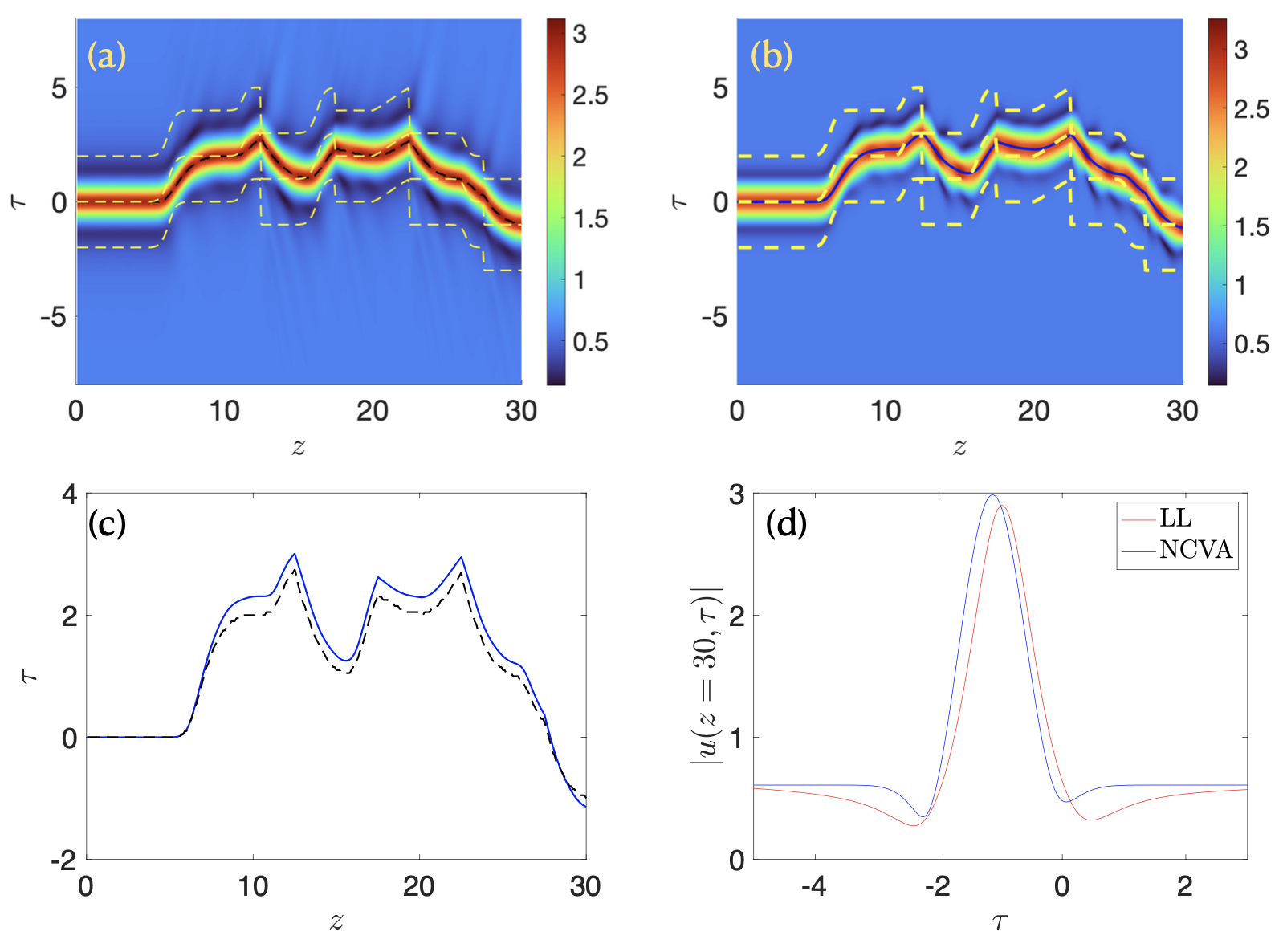}
\caption{Dynamic evolution of a CS using a zigzagging tweezer with 
different degrees of adiabaticity by varying $\tau_f$ and $\beta$ 
(see text for details). These computations were performed by fixing 
$z^{*}=1.25$ [see Eq.~\eqref{tau0}].
(a) The surface plot of the wave amplitude $|u|$ from the LL model. 
We overlay the soliton center $\xi(z)$ (black dashed line) and tweezer 
center $\tau_0(z)$ (central yellow dashed line). We also show 
the maxima of the effective potential $\tau_0(z)\pm \sigma_{\phi}$ 
in yellow dashed lines, depicting the extents of the soliton waveguide. 
(b) Same as (a), but generated from the NCVA ODE calculations. Here the 
NCVA soliton center of mass $\xi(z)$ is shown in blue solid line. 
(c) CS tweezed trajectories for the LL model (black dashed) and NCVA 
reduction (blue solid) that continue to follow the
prescribed trajectory despite the tweezer being moved back-and-forth 
at various speeds. 
(d) CS wave profiles of the LL (red solid) and NCVA (blue solid) at final
time $z=30$, which also demonstrate reasonable quantitative agreement.
This example serves to illustrate how a CS can be robustly manipulated given 
various degrees of adiabaticity.} 
\label{fig:Final}
\end{figure}
%%%%%%%%%% Fig %%%%%%%%%%%%%%%%%%%%%%%%%%%%%%%%%%%%%%

%%%%%%%%%%%%%%%%%%%%%%%%%%%%%%%%%%%%%%%%%%%%%%%%%%%%%%%%%%%%%%%%%
\subsection{Demonstration of a Non-trivial Temporal Tweezing}
 \label{section:FinalExample}
%%%%%%%%%%%%%%%%%%%%%%%%%%%%%%%%%%%%%%%%%%%%%%%%%%%%%%%%%%%%%%%%%

Now that we have studied the parameter regions that give rise to the
different tweezing scenarios, let us demonstrate the robustness of
the CS manipulation when using temporal tweezer parameters $\tau_f$ and 
$\beta$ inside the tweezability region.
In particular, Fig.~\ref{fig:Final} depicts a more general tweezing
scenario where, instead of a single value for the parameters 
$\tau_f$ and $\beta$, we change the parameters at set increments in $z$. 
Specifically, we choose a nontrivial back-and-forth tweezing ``trajectory''
with a zigzagging motion separated in six subintervals
with different degrees of adiabaticity as follows.
For $ 0\leq z<  5$, $\tau_f( 0)=0$, and $\beta =  1$, then 
for $ 5   \leq z< 10$, $\tau_f( 5)=2$, and $\beta =  2$, then 
for $10\leq  z< 15$, $\tau_f(10)=1$, and $\beta = 3$, then 
for $15\leq z< 20$, $\tau_f(15)=2$, and $\beta =  2$, then 
for $20 \leq z<25$, $\tau_f(20)=1$, and $\beta = 0.5$, then 
for $25 \leq z<30$, $\tau_f(25)=-1$, and $\beta = 3$. 
As Fig.~\ref{fig:Final} shows, the CS is successfully tweezed along this
complicated zigzagging orbit (see middle white dashed curve), thus 
showcasing the robustness of the tweezing mechanism
for the full LL dynamics [shown in panel (a)].
Furthermore, as panel (b) shows, the NCVA reduction of the LL system is also
able to predict and reproduce the successful tweezing of the CS along the 
zigzagging orbit with a very good quantitative agreement with the full 
PDE LL calculations. This is also reflected in panel (c) where 
the soliton center $\xi$ is extracted from the LL PDE calculations 
(black dashed  line) together with the variational parameter 
$\xi(z)$ demonstrating good quantitative agreement as well. 
Finally, in panel (d) we depict the LL (red solid line) and NCVA 
(blue solid line) CSs at the end of this complex tweezing scenario,
demonstrating again a good agreement between the original LL PDE model
and its NCVA ODE reduction. 
This tweezing example suggests that, through an appropriate manipulation 
of the phase-modulation of the holding beam, it is possible to trap and 
move the CS at will, based on the proposed tweezing mechanism and for a
(revealed herein) wide interval of the associated dynamical parameters.
Furthermore, examining the dynamics on a soliton manifold (NCVA calculations) 
yields valuable insights into the full dynamics, while enabling a 
computational platform that is significantly cheaper.

%%%%%%%%%%%%%%%%%%%%%%%%%%%%%%%%%%%%%%%%%%%%%%%%%%%%%%%%%%%%%%%%%
\section{Conclusions and Future Challenges}
\label{secConclusion}
%%%%%%%%%%%%%%%%%%%%%%%%%%%%%%%%%%%%%%%%%%%%%%%%%%%%%%%%%%%%%%%%%

In Ref.~\cite{tweeze} the authors showed experimentally that
a cavity soliton (CS) stored in a passive loop of an
optical fiber could be temporally tweezed by manipulating the
phase-modulation of the holding beam.
Motivated by this experimental work, we studied herein the possibility of 
tweezing more generally CSs using various profiles of the holding beam and in 
particular varying the degree of adiabaticity used in the tweezing ``trajectory''.
We modeled the system using a variant of the complex Ginzburg-Landau
(cGL) equation in the form of the Lugiato-Lefever (LL) partial differential 
equation with additional terms accounting for the incorporation of the
tweezer from the phase-modulation of the holding beam.

In our study, we assume a Gaussian phase-modulation that gives
rise to an effective tweezer potential, related 
to the derivative of the phase-modulation profile, that takes the 
shape of a localized trapping profile.
We find that, depending on the tweezer parameters, the following two 
different outcomes are possible.
(a) For sufficiently slow (adiabatic) tweezing motion, the tweezer is generically 
capable of transporting the CS to any desired location.
(b) On the other hand, for relatively large tweezer speeds the CS slides out
of the tweezing potential and is left behind to exist in the bulk of the medium
while the tweezer moves away without carrying along the CS.
As part of our theoretical study, we develop a Lagrangian formalism for the 
modified the LL equation that intrinsically includes gain and loss terms. 
To tackle this out-of-equilibrium, non-Hamiltonian system we use the
recently developed non-conservative variational approximation 
(NCVA) of Ref.~\cite{JuliaNCVA}, based on the earlier formulation
of Refs.~\cite{Galley,Galley:14}. We use the NCVA approach with an
appropriately crafted CS ansatz to reduce the original LL dynamics
(a partial differential equation) to a set of coupled ordinary differential
equations on the ansatz parameters (height, position, width, phase, 
velocity, and chirp).
We show how the NCVA reduction approach is able to qualitatively, and to a 
significant extent quantitatively, describe the different tweezing scenarios.
In particular, we notice that the NCVA is capable of predicting quite 
accurately the threshold in the tweezer parameter space.

Our analysis of not only static, but also dynamical tweezing offers 
insights into the design of localized tweezers used in optical information 
processing, in turn enabling the potential trapping of ultrashort pulses 
of light and dynamical moving of those around in time, with respect to, 
and independently of other pulses of light.
Furthermore, from our study of the LL equation and the NCVA approach, we 
have developed a process to identify regions of tweezability which can aid 
toward the experimental design and identification of regimes of reliable
temporal tweezing used for information processing.

A possible extension of this work pertains the case of a periodic
modulation of the holding beam that induces a periodic effective
tweezer potential. It would be interesting to study in detail the 
possible manipulation of the CSs using such periodic (or even 
quasi-periodic) modulation as it was studied in Refs.~\cite{rcg25,rcg34} 
for the (otherwise conservative) nonlinear Schr\"odinger case.
Another avenue of potential interest would be the study of tweezability
of not only cavity solitons but also vortices in two-dimensional settings 
in the presence of gain and loss similar to what has been
reported in the conservative case of the nonlinear Schr\"odinger
equation; see Refs.~\cite{rcg65,rcg108} and references therein.
In general, the extension of considerations presented herein
for one spatial dimension to the context of higher-dimensional
settings would be a particularly interesting direction for future work.

In reference to temporal tweezing, multiple CSs can be present 
simultaneously and independently at arbitrary locations in a passive 
loop of optical fiber. Therefore, a relevant extension would 
be to add interactions of multiple CSs in the system. 
Indeed, a systematic study of the nonlinear states in the present
setting and of their properties would be of interest in its own right.
In addition, investigating 
the dynamics, interactions, and tweezability of the CS by allowing 
for long-range soliton interactions is necessary to understand an 
effective treatment of a CS, each of which constitute an ideal bit 
in optical information processing.
It is relevant to mention in that regard that the recent review
of Ref.~\cite{erki2} has summarized the extensive control that
this framework enables both in the way of phase, as well as
of intensity of the inhomogeneous driving field. Additionally,
recent experiments of pulsed driving are also discussed therein
towards the direction of producing flexible and efficient optical
frequency combs. 
Another possible extension of the temporal tweezing study is to 
analyze the linearization spectrum of the system and identify the 
(in)stability transitions between the different states we 
identified. 
This type of analysis can be extended from statically tweezed states
to potential traveling ones.
Indeed, it would be interesting to identify the co-traveling states
in the case of a traveling tweezer and to identify whether
the transition, e.g., from tweezing to disappearance of the
pulse can be identified with a bifurcation phenomenon.
Such directions are currently in progress and will be reported
in future publications.

%%%%%%%%%%%%%%%%%%%%%%%%%%%%%%%%%%%%%%%%
\begin{appendices}

\section{NCVA System of Equations for Temporal Tweezing of Cavity Soliton}
\label{AppendixA}
%%%%%%%%%%%%%%%%%%%%%%%%%%%%%%%%%%%%%%%%%%%%%%%%%%%%%%%%%%%%%%%%%%%%%%%%%%%

In this Appendix, we present the resulting modified Euler-Lagrange equations of motion from the LL Eq.~(\ref{eq:LLETweeze}) for temporal tweezing based on the NCVA where the over-dot denotes derivative with respect to $z$.
Before presenting these equations, we provide the effective Lagrangian $\overline{L}$ evaluated on the Gaussian ansatz Eq.~\eqref{6pAnsatzTweeze}:

{\fontsize{6}{0}\selectfont
\def\myskip{-1.2ex}{
\begin{align}
    \overline{L}&={\sqrt{\pi}(-c\dot \xi+\dot b)a^2\sigma+\frac{\sqrt{\pi}a^2\sigma^3}{2}\dot d+\frac{\sqrt{\pi}a^2}{2\sigma}+2\sqrt{\pi}a^2d^2\sigma^3+\sqrt{\pi}a^2c^2\sigma-\frac{\sqrt{2\pi}a^4\sigma}{4}}\\\nonumber&{+\Delta \sqrt{\pi}a^2\sigma+\frac{\sqrt{\pi}{\rm e}^{-\frac{\tau_0^2-2\tau_0\xi+\xi^2}{\sigma_{\phi}^2+\sigma^2}}h_{\phi}^2\sigma_{\phi}\sigma(\tau_0-\xi)^2a^2}{(\sigma^2+\sigma_{\phi}^2)^{5/2}}+\frac{\sqrt{\pi}{\rm e}^{-\frac{\tau_0^2-2\tau_0\xi+\xi^2}{\sigma_{\phi}^2+\sigma^2}}h_{\phi}^2a^2\sigma^3}{2\sigma_{\phi}(\sigma_{\phi}^2+\sigma^2)^{3/2}}}\\\nonumber&{-2\sqrt{\pi}|v_s|^2a^2\sigma+\frac{4\sqrt{2\pi}{\rm e}^{-\frac{\tau_0^2-2\tau_0\xi+\xi^2}{2\sigma_{\phi}^2+\sigma^2}}h_{\phi}\sigma_{\phi}\sigma (\tau_0-\xi)a^2\left(2d\sigma^2\tau_0-2d\sigma^2\xi+2\sigma_{\phi}^2c+c\sigma^2\right)}{(2\sigma_{\phi}^2+\sigma^2)^{5/2}}}\\\nonumber&{-\frac{4\sqrt{2\pi}{\rm e}^{-\frac{\tau_0^2-2\tau_0\xi+\xi^2}{2\sigma_{\phi}^2+\sigma^2}}h_{\phi}\sigma_{\phi}\sigma^3 a^2 d}{(2\sigma_{\phi}^2+\sigma^2)^{3/2}}.}
\end{align}}
 }
\noindent
Substituting the expression for the effective Lagrangian into the 
modified Euler-Lagrange equations Eqs.~\eqref{NCVAODE} yields the 
following  explicit ODE system (obtained using Maple):
{\fontsize{6}{0}\selectfont
\def\myskip{-1.2ex}
\begin{eqnarray}
   { \dot{a}}&=&{\frac{-2}{\sqrt{\pi}a\sigma^3(2\sigma_{\phi}^2+\sigma^2)^{5/2}}\Bigg[-\sqrt{2\pi}\sigma_{\phi}h_{\phi}\sigma^3a^2{\rm e}^{-\frac{(\xi-\tau_0)^2}{2\sigma_{\phi}^2+\sigma^2}}\left(\sigma^2-2(\xi-\tau_0+\sigma_{\phi})(\xi-\tau_0-\sigma_{\phi})\right)}\nonumber\\[\myskip]&&{+(2\sigma_{\phi}^2+\sigma^2)^{5/2}\left(\sqrt{\pi}a^2d\sigma^3-\frac{3\sigma^2}{8}I_b+\frac{I_d}{4}\right)\Bigg]}
\end{eqnarray}

%\begin{eqnarray}
 %   \dot{b}&=&\frac{-2}{\sqrt{\pi}a\sigma^3(2\sigma_{\phi}^2+\sigma^2)^{5/2}}\Bigg[-\sqrt{2\pi}\sigma_{\phi}h_{\phi}\sigma^3a^2{\rm e}^{-\frac{(\xi-\tau_0)^2}{2\sigma_{\phi}^2+\sigma^2}}\left(\sigma^2-2(\xi-\tau_0+\sigma_{\phi})^2\right)\nonumber\\[\myskip]&&+(2\sigma_{\phi}^2+\sigma^2)^{5/2}\left(\sqrt{\pi}a^2d\sigma^3+\frac{3\sigma^2}{8}I_b-\frac{I_d}{4}\right)\Bigg]
%\end{eqnarray}
\begin{eqnarray}
    {\dot{b}} &=& {-\frac{3}{4\sqrt{\pi}a^2\sigma^2\sigma_{\phi}(2\sigma_{\phi}^2+\sigma^2)^{9/2}(\sigma_{\phi}^2+\sigma^2)^{9/2}}\times} \nonumber\\[\myskip] && {\Bigg[-\frac{5\sigma_{\phi}(2\sigma_{\phi}^2+\sigma^2)^{9/2}(\sigma_{\phi}^2+\sigma^2)^{9/2}}{6}\bigg(a^2\sqrt{\pi}\left(-\frac{8}{5}+\sqrt{2}a^2\sigma^2+\frac{8}{5}(c^2+2|v_s|^2-\Delta)\sigma^2\right)-\frac{6\sigma}{5}(aI_a-\frac{4cI_c}{3}}\nonumber\\[\myskip] &&{-\frac{2\sigma I_s}{3})\bigg)-8\sqrt{2\pi}h_{\phi}\sigma_{\phi}^2a^2\sigma^4(\sigma_{\phi}^2+\sigma^2)^{9/2}{\rm e}^{-\frac{(\xi-\tau_0)^2}{\sigma^2+2\sigma_{\phi}^2}}\bigg(\delta_1\sigma^6+\delta_2\sigma^4+\delta_3\sigma^2+\delta_4\bigg)}\nonumber\\[\myskip] &&{+\sqrt{\pi}h_{\phi}^2a^2\sigma^2(2\sigma_{\phi}^2+\sigma^2)^{9/2}{\rm e}^{-\frac{(\xi-\tau_0)^2}{\sigma_{\phi}^2+\sigma^2}}\bigg(\sigma^8+\delta_5\sigma^6+\delta_6\sigma^4+\delta_7\sigma^2+\delta_8\bigg)\Bigg]}
\end{eqnarray}
%%%%%
\begin{eqnarray}
    {\dot{c}} &=& {\frac{-24}{\sqrt{\pi}a^2\sigma\sigma_{\phi}(\sigma_{\phi}^2+\sigma^2)^{7/2}(2\sigma_{\phi}^2+\sigma^2)^{7/2}}\Bigg[\sqrt{2\pi}(\sigma_{\phi}^2+\sigma^2)^{7/2}\sigma_{\phi}^2\sigma h_{\phi}a^2{\rm e}^{-\frac{(\xi-\tau_0)^2}{2\sigma_{\phi}^2+\sigma^2}}}\times\nonumber\\[\myskip]&&{\bigg(\left(d\xi-d\tau_0-\frac{c}{6}\right)\sigma^4+\left(-\frac{2d\xi^3}{3}+(2d\tau_0+\frac{c}{3})\xi^2+2(-\frac{c\tau_0}{3}+(\sigma_{\phi}^2-\tau_0^2)d)\xi+\frac{(\tau_0^2-2\sigma_{\phi}^2)c}{3}+2d\tau_0(\tau_0^2-3\sigma_{\phi}^2)\right)\sigma^2}\nonumber\\[\myskip]&&{+\frac{2\sigma_{\phi}^2c(\xi-\tau_0+\sigma_{\phi})(\xi-\tau_0-\sigma_{\phi})}{3}\bigg)-\frac{(2\sigma_{\phi}^2+\sigma^2)^{7/2}}{24}\bigg(-\left(\sigma_{\phi}^2+\sigma^2\right)^{7/2}\sigma_{\phi}(I_b c+I_{\xi})}\nonumber\\[\myskip]&&{+(\xi-\tau_0)\sigma \sqrt{\pi}h_{\phi}^2a^2{\rm e}^{-\frac{(\xi-\tau_0)^2}{\sigma_{\phi}^2+\sigma^2}}\left(\sigma^4-\sigma_{\phi}^2\sigma^2+2\sigma_{\phi}^2(\xi-\tau_0+\sigma_{\phi})(\xi-\tau_0-\sigma_{\phi})\right)\bigg)\Bigg]}  
\end{eqnarray}
%%%%%
\begin{eqnarray}
    {\dot {d}} &=&{-\frac{1}{4\sqrt{\pi}a^2\sigma^4}\bigg(a^2\sqrt{\pi}(16d^2\sigma^4+\sqrt{2}a^2\sigma^2-4)-2aI_a\sigma+4I_{\sigma}\sigma^2\bigg)}\\\nonumber&&{-\frac{4}{\sqrt{\pi}a^2\sigma^4\sigma_{\phi}(\sigma_{\phi}^2+\sigma^2)^{9/2}(2\sigma_{\phi}^2+\sigma^2)^{9/2}}\Bigg[\sqrt{2\pi}(\sigma_{\phi}^2+\sigma^2)^{9/2}\sigma_{\phi}^2\sigma^4 h_{\phi}a^2 {\rm e}^{-\frac{(\xi-\tau_0)^2}{2\sigma_{\phi}^2+\sigma^2}}\times} \nonumber\\[\myskip]&&{\left(d\sigma^6+\delta_9\sigma^4+\delta_{10}\sigma^2+\delta_{11}\right)-\frac{(2\sigma_{\phi}^2+\sigma^2)^{9/2}}{8}a^2\sqrt{\pi}h_{\phi}^2{\rm e}^{-\frac{(\xi-\tau_0)^2}{\sigma_{\phi}^2+\sigma^2}} \sigma^4\left(\sigma^6+\delta_{12}\sigma^4+\delta_{13}\sigma^2+\delta_{14}\right)\Bigg]}
\end{eqnarray}

\begin{eqnarray}
{    \dot{\sigma}}&=& {\frac{4}{\sqrt{\pi}a^2\sigma^2(2\sigma_{\phi}^2+\sigma^2)^{5/2}}\Bigg[-\sqrt{2\pi}\sigma_{\phi}h_{\phi}\sigma^3a^2{\rm e}^{-\frac{(\xi-\tau_0)^2}{2\sigma_{\phi}^2+\sigma^2}}\left(\sigma^2-2(\xi-\tau_0+\sigma_{\phi})(\xi-\tau_0-\sigma_{\phi})\right)}\nonumber\\[\myskip]&&{+(2\sigma_{\phi}^2+\sigma^2)^{5/2}\left(\sqrt{\pi}a^2d\sigma^3-\frac{\sigma^2I_b}{8}+\frac{I_d}{4}\right)\Bigg]}
\end{eqnarray}

%%%%
\begin{eqnarray}
\label{ODE-xi}
   { \dot{\xi}} &&={\frac{-4}{\sqrt{\pi}(2\sigma_{\phi}^2+\sigma^2)^{3/2}a^2\sigma}\Bigg[\sqrt{2\pi}\sigma_{\phi}h_{\phi}a^2\sigma(\xi-\tau_0){\rm e}^{-\frac{(\xi-\tau_0)^2}{2\sigma_{\phi}^2+\sigma^2}}}\nonumber\\[\myskip]&&{-\sqrt{2\sigma_{\phi}^2+\sigma^2}\left(\sqrt{\pi}c\sigma a^2+\frac{I_c}{2}\right)(\sigma_{\phi}^2+\frac{\sigma^2}{2})\Bigg]}
\end{eqnarray}

} %\fontsize{6}{0}\selectfont

\noindent
where
\begin{eqnarray}
\delta_1&=& d, 
\nonumber
\\
\nonumber
\delta_2&=& -4d\xi^2+(8d\tau_0+c)\xi-c\tau_0+4d(\sigma_{\phi}^2-\tau_0^2),
\nonumber
\\
\nonumber
\delta_3&=& 2\bigg[\frac{2d}{3}\xi^4+\frac{(-8d\tau_0-c)}{3}\xi^3+\left(c\tau_0+4d(\tau_0^2-\sigma_{\phi}^2)\right)\xi^2\nonumber\\&+&\bigg(\left(2\sigma_{\phi}^2-\tau_0^2\right)c-\frac{8d\tau_0}{3}\left(\tau_0^2-3\sigma_{\phi}^2\right)\bigg)\xi\nonumber\\&+&\left(-2\tau_0\sigma_{\phi}^2+\frac{1}{3}\tau_0^3\right)c+\frac{2d}{3}\left(3\sigma_{\phi}^4-6\tau_0^2\sigma_{\phi}^2+\tau_0^4\right)\bigg],
\nonumber
\\
\nonumber
\delta_4&=& -\frac{4c\sigma_{\phi}^2}{3}(\xi-\tau_0)\left((\xi-\tau_0)^2-3\sigma_{\phi}^2\right),
\\
\delta_5&=& 2\left(-\frac{\xi^2}{3}-\frac{\tau_0^2}{3}+\sigma_{\phi}^2+\frac{2\tau_0\xi}{3}\right), 
\nonumber
\\
\nonumber
\delta_6&=& 4\sigma_{\phi}^2\left(\xi^2-2\tau_0\xi+\tau_0^2+\frac{\sigma_{\phi}^2}{4}\right),
\nonumber
\\
\nonumber
\delta_7&=& -\frac{4\sigma_{\phi}^2}{3}(\xi-\tau_0)^2\left((\xi-\tau_0)^2-\frac{9\sigma_{\phi}^2}{2}\right),
\nonumber
\\
\nonumber
\delta_8&=& \frac{4\sigma_{\phi}^6}{3}(\xi-\tau_0)^2,
\nonumber
\\
\delta_9&=& -8(\xi-\tau_0)(d\xi-d\tau_0-\frac{3c}{8}),
\nonumber
\\
\delta_{10}&=& 2\Bigg(2d\xi^4+(-8d\tau_0-c)\xi^3+\left(-4d(\sigma_{\phi}^2-3\tau_0^2)+3c\tau_0\right)\xi^2\nonumber\\&+&\left(8\tau_0d(\sigma_{\phi}^2-\tau_0^2)-3c(\tau_0^2-2\sigma_{\phi}^2)\right)\xi\nonumber\\&+&2d\left(-3\sigma_{\phi}^4-2\tau_0^2\sigma_{\phi}^2+\tau_0^4\right)+\tau_0c (\tau_0^2-6\sigma_{\phi}^2)\Bigg),
\nonumber
\\
\delta_{11}&=&16\sigma_{\phi}^2\Bigg[-\frac{c}{4}\xi^3+\left(d\sigma_{\phi}^2+\frac{3c\tau_0}{4}\right)\xi^2+\left(-2d\tau_0\sigma_{\phi}^2+\frac{3c}{4}(\sigma_{\phi}^2-\tau_0^2)\right)\xi\nonumber\\&+&\sigma_{\phi}^2(\tau_0^2-\sigma_{\phi}^2)d+\frac{\tau_0 c}{4}(\tau_0^2-3\sigma_{\phi}^2)\Bigg],
\nonumber
\\
\delta_{12}&=& -2(\xi-\tau_0)^2,
\nonumber
\\
\delta_{13}&=& 8\sigma_{\phi}^2\left((\xi-\tau_0)^2-\frac{3}{8}\sigma_{\phi}^2\right),
\nonumber
\\
\delta_{14}&=& -4\sigma_{\phi}^2\bigg(\xi^4-4\tau_0\xi^3+\left(6\tau_0^2-\frac{5\sigma_{\phi}^2}{2}\right)\xi^2+\left(5\tau_0\sigma_{\phi}^2-4\tau_0^3\right)\xi \nonumber\\&+&\tau_0^4-\frac{5\sigma_{\phi}^2}{2}\tau_0^2+\frac{\sigma_{\phi}^4}{2}\bigg).
\end{eqnarray}

%input{AppendixB}
%%%%%%%%%%%%%%%%%%%%%%%%%%%%%%%%%%%%%%%%%%%%%%%%%%%%%%%%%%%%%%%

\end{appendices}
%%%%%%%%%%%%%%%%%%%%%%%%%%%%%%%%%%%%%%%%

%%%%%%%%%%%%%%%%%%%%%%%%%%%%%%%%%%%%%%%%
%%%%%%%%%%%%%%%%%%%%%%%%%%%%%%%%%%%%%%%%
\section*{Declarations}

\begin{itemize}

\item
{\bf Ethics approval and consent to participate:}
Not applicable

\item
{\bf Consent for publication:}
Not applicable

\item
{\bf Availability of data and materials:}
Not applicable

\item
{\bf Competing interests:}
The authors have no relevant financial or non-financial interests to disclose.
The authors have no competing interests to declare that are relevant to the content of this article.

\item
{\bf Funding:}
RCG acknowledges the support from NSF-1603058 and NSF-2110038.
PGK acknowledges the support from NSF-2204702 and NSF-2110030.

\item
{\bf Authors' contributions:}
All authors contributed equally.

\item
{\bf Acknowledgements}:
Not applicable

\end{itemize}

%\bibliography{references4}% common bib file
%%% if required, the content of .bbl file can be included here once bbl is generated
%%%\input sn-article.bbl


\end{document}